\documentclass[twoside,submission,publicdomain,11pt]{eptcs}

\newcommand{\pcindextermi}[1]{\textbf{#1}}

\usepackage{amssymb}
\usepackage{calc}
\usepackage{ifthen}
\usepackage{epsfig}
\usepackage{varioref}
\usepackage{multicol}
\usepackage{color} 
\usepackage{daad}
\usepackage{daad-bbbb}
\usepackage{daad-color}

\usepackage{rslenv}
\usepackage{psboxit}
\usepackage{boitesexamples}

\newcommand{\eod}{.}

\newcommand{\nref}[1]{}

\newcommand{\nyl}{\pos{}{\\}}
\newcommand{\pos}[2]{#1}

\newcommand{\boks}[2]{\vspace*{2mm} 
                      \enumerer\boiteepaisseavecuntitre{\brcolor{Schema:} #1}
                      #2
                      \endboiteepaisseavecuntitre\senumerer}
        
\newcounter{aircraft}
\newcommand{\nac}{\stepcounter{aircraft}}

\renewcommand{\nrslframebox}[2]{\vspace*{2mm} 
                               \boiteepaisseavecuntitre{Aircraft
                               \bbcolor{Example \nac\arabic{aircraft}:} #1}
                               #2
                               \endboiteepaisseavecuntitre}

\newcommand{\unrslframebox}[2]{\vspace*{2mm} 
                               \boiteepaisseavecuntitre{Aircraft
                               \bbcolor{Example \arabic{aircraft}:} #1}
                               #2
                               \endboiteepaisseavecuntitre}

\newcommand{\enumerer}{
   \renewcommand{\begynd}{\begin{itemize}}
   \renewcommand{\pind}{\item }
   \renewcommand{\afslut}{\end{itemize}}}

\newcommand{\senumerer}{
   \renewcommand{\begynd}{}
   \renewcommand{\pind}{}
   \renewcommand{\afslut}{}}

\newcommand{\begynd}{}
\newcommand{\pind}{}
\newcommand{\afslut}{}

\newcommand{\todaytime}{\today: 
  \ifthenelse{\value{dbhours}<10}{0}{}\thedbhours:%
  \ifthenelse{\value{dbmins}<10}{0}{}\thedbmins\,\ifthenelse{\value{dbhours}<12}{am}{}}

\renewcommand{\pt}[1]{#1}

\renewcommand{\psno}{}
\renewcommand{\mpsno}{}

\title{\sf\brcolor{\sf Domain Analysis \& Description}\\[1mm]
  \sf\bbcolor{\Large\sf The Implicit and Explicit
  Semantics Problem}}
\author{\sf\brcolor{\sf Dines Bj{\o}rner}\\[2mm] 
 \sf\bbcolor{\sf Fredsvej 11, DK-2840 Holte, Danmark}\\
 \sf\bbcolor{\sf Technical University of Denmark, DK-2800 Kgs.Lyngby,
  Denmark}\\ 
 \sf\bbcolor{\sf E--Mail: bjorner@gmail.com, URL: www.imm.dtu.dk/\~{}db}}
\date{\sf\bgcolor{\sf \today}}

\newcommand{\titlerunning}{Domain Analysis \& Description}
\newcommand{\authorrunning}{Dines Bj{\o}rner}

\newcommand{\xdbevenfoot}%
{$\overline{\mbox{\tiny\sf{\copyright\ D.Bj{\o}rner}\ 
    2018, Fredsvej 11, 2840 Holtes, Denmark.\today: 
       \ifthenelse{\value{dbhours}<10}{0}{}\thedbhours:%
        \ifthenelse{\value{dbmins}<10}{0}{}\thedbmins\rm}}$ 
   \hfil \thepage \hfil $\overline{\mbox{\brcolor{\tiny\sf The
   Manifest Domain Analysis \& Description Approach}}}$} 
 
\newcommand{\xdboddfoot}%
{$\overline{\mbox{\bbcolor{\tiny\sf to Implicit and Explicit
  Semantics}}}$     
   \hfil \thepage \hfil $\overline{\mbox{\tiny\sf \copyright\
   {\sf D.Bj{\o}rner 2018. Fredsvej 11, 2840 Holtes, Denmark. \today: 
       \ifthenelse{\value{dbhours}<10}{0}{}\thedbhours:%
        \ifthenelse{\value{dbmins}<10}{0}{}\thedbmins\rm}}$}}

\newcommand{\faoce}{}

\newcommand{\bnovchg}{}
\newcommand{\enovchg}{}

\newcommand{\pdindextermi}[1]{#1}

\newcommand{\pdindextermiii}[3]{#1 #2 #3}

\renewcommand{\nyl}{}

\begin{document} \HHHH

\dbcalctime

\maketitle\HHHH

\mnewfoil

\pos{\begin{abstract}}{\centerline{\brcolor{Summary}}}\HHHH

\noindent
\begynd
\pind The domain analysis \& description calculi introduced in
      \cite{BjornerFAoC2015MDAAD}  
\begynd
\pind is shown to alleviate the issue of implicit semantics \cite{impex-0,impex-1}.
\pind The claim is made that domain descriptions, 
\begynd
\pind whether informal, or as also here, formal, 
\pind amount to an explicit semantics \nyl for what is
      otherwise implicit if not described\,!
\afslut
\afslut
\mnewfoil
\pind I claim that \cite{BjornerFAoC2015MDAAD}  provides \nyl
      an answer to the claim in both \cite{impex-0,impex-1} that
\begynd
\pind ``The contexts of the systems in these cases
      are treated as second-class citizens\ \ldots'',
\pind respectively
\pind ``In general, modeling languages are not equipped with
      resources, concepts or entities 
      handling explicitly domain engineering features and
      characteristics (domain knowledge) 
      in which the modeled systems evolve''.
\afslut
\afslut
\pos{\end{abstract}}{}

\mnewfoil


\nbbbbb{Introduction}

\bbbb{On the Issues of Implicit and Explicit Semantics}

\begynd
\pind In \cite{impex-0} the issues of implicit and explicit semantics 
      are analysed.
\pind It appears, from \cite{impex-0}, that when an issue
\begynd
\pind of software requirements or 
\pind of the context, or, as we shall call it, the domain, 
\pind is not prescribed or described 
\pind to the extent that is relied upon in the software design, 
\pind then it is referred to as an issue of implicit semantics. 
\afslut
\pind Once prescribed, respectively described, \nyl that issue becomes one of
      explicit semantics. 
\pind In this \pos{paper we}{invited talk I} offer \nyl
      \brcolor{a calculus for analysing \& describing domains
        \pos{(i.e., contexts),}{} 
      \nyl a calculus that allows you \nyl to systematically and formally
      describe domains}.
\afslut

\nbbbb{A Triptych of Software Engineering}

\begynd
\pind The dogma is:
\begin{itemize}
\item \sfsl{before \bbcolor{software}
      can be \bbcolor{designed}}
\begynd
\pind \sfsl{we must understand its \bbcolor{requirements};}
\afslut 
\item and \sfsl{before
      we can \bbcolor{prescribe} the \bbcolor{requirements}}
\begynd
\pind \sfsl{we must understand the \bbcolor{domain,}}
\pind \sfsl{that is, \bbcolor{describe} the domain.}
\afslut 
\end{itemize}
\afslut

\mnewfoil
\begynd
\pind A strict, but not a necessary, interpretation of this dogma thus
      suggests that software development ``ideally'' proceeds in three
      phases:
\begin{itemize}
\item First a phase of \brcolor{\textsf{\textsl{domain engineering}}} in which
      an analysis of the application domain leads to a description
      of that domain.\footnote{\LLLL This phase is often misunderstood. On
      one hand we expect domain stakeholders, e,g,, \sfsl{bank} associations and
      university economics departments, to establish ``a family'' of \sfsl{bank}
      domain descriptions: taught when training and educating new
      employees, resp.\ students. Together this 'family' covers as
      much as is known about \sfsl{banking}. On the other hand we
      expect each new 
      \sfsl{bank} application (software) development to ``carve'' out a
      ``sufficiently large'' 
      description of the domain it is to focus on. Please replace the
      term \sfsl{bank} with an appropriate term for the domain for which
      You are to develop software.}
\item Then a phase of \brcolor{\textsf{\textsl{requirements
      engineering}}} in which 
      an analysis of the domain description leads to a prescription of
      requirements to software for that domain.
\item And, finally, a phase of \brcolor{\textsf{\textsl{software design}}} in
      which an analysis of the requirements prescription leads to software
      for that domain.
    \end{itemize}
\afslut
\mnewfoil

\begynd
\pind Proof of program, i.e., software code, correctness can be
      expressed as:
\begin{itemize}
\item \brcolor{$\mathcal{D}, \mathcal{S} \models \mathcal{R}$}
\end{itemize}
\pind which we read as:
\begynd
\pind proofs that \brcolor{$\mathcal{S}$}oftware
\pind is correct with respect to \brcolor{$\mathcal{R}$}equirements
\pind implies references to the \brcolor{$\mathcal{D}$}omain.
\afslut
\afslut

\nbbbb{Contexts \cite{impex-0} {\IS} Domains \cite{BjornerFAoC2015MDAAD}}

\begynd
\pind Often the domain is referred to as the \bbcolor{context}.
\pind We treat contexts, i.e., domain descriptions \nyl as first class
      citizens \nyl \cite[Abstract, Page 1, lines 9--10]{impex-0}. 
\pind By emphasizing the formalisation of domain descriptions \nyl
      we thus focus on the \sfsl{explicit} semantics.  
\pind Our approach, \cite{BjornerFAoC2015MDAAD}, summarised in
      Sect.\,\pos{\ref{impex-daad}}{2.} of this paper, \nyl
      thus represents a formal approach to \nyl the description of contexts
      (i.e., domains) \nyl \cite[Abstract, Page 1, line 12]{impex-0}.
\pind By a \brcolor{domain}, i.e., a context, \brcolor{description}, we shall here
      understand an \textbf{explicit semantics} of what is usually not
      specified and, when not so,  referred to as \bbcolor{implicit
semantics}\footnote{\LLll ``The contexts  \ldots\
are treated as second-class citizens: in general, the modelling is implicit
and usually distributed between the requirements model and the system
model.'' \LLll \cite[Abstract, Page 1, lines 9--12]{impex-0}.}.
\afslut

\nbbbb{Semantics}

\begynd
\pind I use the term `semantics' rather than the term `knowledge'.
\pind The reason is this:
\begynd
\pind The entities are what we can meaningfully speak about. \LLLL\sf
\begynd
\pind That is, the names of the endurants and perdurants,
\pind of their being atomic or composite, discrete or continuous,
\pind parts, components or materials,
\pind their unique identifications, mereologies and attributes,
\pind and the types, values and use of operations over these,
\afslut \rm\HHHH
      form the language spoken by practitioners in the domain.
\afslut
\pind It is this language 
\begynd
\pind its base syntactic quantities and
\pind semantic domains 
\afslut we structure and ascribe a semantics.
\afslut

\nbbbb{Method \& Methodology}

\begynd
\pind By a \brcolor{method} I understand
\begynd
\pind a set of principles 
\pind for selecting and applying 
\pind techniques and tools
\afslut for constructing a manifest or an
        abstract artifact.
\pind By \brcolor{methodology} I understand the study and knowledge of
      methods. 
\pind \textbf{My work is almost exclusively in the area of
      methods and methodology.} 
\afslut

\nbbbb{Computer \& Computing Sciences}

\begynd
\pind By \brcolor{computer science} I understand
\begynd
\pind the study and knowledge about the things
\pind  that can exist inside computing devices.
\afslut
\afslut

\begynd
\pind By \brcolor{computing science} I understand
\begynd
\pind the study and knowledge about how to construct the things
\pind that can exist inside computing devices.
\afslut Computing science is also often referred to \nyl as
        \sfsl{programming methodology.}
\pind \textbf{My work is almost exclusively in the area of
      computing science.}
\afslut

\nbbbb{Software and Systems Engineering}

\begynd
\pind By \brcolor{software engineering} I understand the triplet of
\begynd
\pind \bbcolor{domain engineering},
\pind \bbcolor{requirements engineering}, and
\pind \bbcolor{software design}.
\afslut
\pind \textbf{My work has almost exclusively been in the area of
  methodologies for large scale software -- beginning with
  compilers (\texttt{CHILL} and \texttt{Ada},
  \cite{Haff87,vdm:CCITT80,e:db:Bj80f,Clem84,Oest86}).} 
\afslut

\nbbbbb{The Analysis \& Description Prompts}\label{impex-daad}



\begynd
\pind We present a calculus of analysis and description
      prompts\footnote{\LLLL\sf Prompt, as a verb: to move or induce
      to action; to occasion or incite; inspire;  to assist (a person
      speaking) by \textsl{''suggesting something to be said''}.}.
\pind The presentation here is a very short\pos{, 12 pages,}{} version of
      \cite[Sects.\,2--4, 31 pages]{BjornerFAoC2015MDAAD}.
\begynd
\pind These prompts are tools \nyl that the
      domain analyser \& describer uses.
\pind The domain analyser \& describer is
      in the domain, \nyl sees it, can touch it,  and then applies the 
      prompts, \nyl in some orderly fashion, to what is being
      observed. 
\begynd
\pind So, on one hand, there is the necessarily informal domain, and, 
\pind on the other hand, there are the seemingly formal prompts 
\pind and the \textsl{``suggestions for something to be said''}, i.e.,
      written down: narrated and formalised. \pos{See Fig.\,\vref{impex.ontology}.}{} 
\afslut
\afslut
\afslut
\mnewfoil
\vDBfigure{ontology}{\pos{15.4}{15.9}cm}{An Ontology for
  Manifest Domains}{impex.ontology}
\mnewfoil
\noindent
\begynd
\pind The figure suggests a number of \bbcolor{analysis} and
      \brcolor{description} prompts.
\begynd
\pind The domain analyser \& describer is ``positioned'' at the
      top\pos{, the ``root''}{}. 
\pind If what is observed can be conceived and
      described \nyl then it \bbcolor{is an \sfsl{entity}}.
\pind If it can be described
      as a ``complete thing'' \nyl at no matter which given snapshot 
      of time \nyl then it \bbcolor{is an \sfsl{endurant}}.
\pind If it is an entity but for which
      only a fragment exists \nyl if we look at 
      or touch them at any given snapshot in time, \nyl then it
      \bbcolor{is a \sfsl{perdurant}}.
\afslut
\afslut
\mnewfoil

\begynd
\pind The concepts of \sfsl{endurants} and \sfsl{perdurants}
      may seem novel to some readers. So we elaborate a bit.
\begynd
\pind First we must recall that we are trying to describe
      aspects a real worlds.
\pind That is, to model, in narrative and in formal terms,
      what has puzzled philosophers since antiquity.
\pind One can therefore not expect to define the terms
      \sfsl{`endurants'} and \sfsl{`perdurants'} as one define terms
      in computer science and mathematics. 
\afslut
\pind Here, then are some ``definitions'', i.e. some delineations,
      some ``encirclings'' of crucial concepts.
\afslut

\bookdefn{Entity}{
\begynd
\pind By an \pdindextermi{entity} we shall
      understand a \pdindextermi{phenomenon}, i.e., something
\begynd
\pind that can be \pcindextermi{observe}d, i.e., be
\begynd  
\pind seen or touched by humans,
\pind \sfsl{or} that can be \pcindextermi{conceive}d 
\pind as an \pcindextermi{abstraction}  of an entity;
\afslut 
\pind alternatively,
\begynd
\pind a phenomenon is an entity, \sfsl{if it exists, it is
      \pdindextermi{``being''}, 
\pind it is that which makes a \pcindextermi{``thing''} what it is: \nyl
      essence, essential nature}\pos{ \cite[Vol.\,I, pg.\,665]{OED}}{}\eod
\afslut
\afslut
\afslut
}
\mnewfoil

\bookdefn{Endurant}{%
\begynd
\pind By an \pdindextermi{endurant} we shall
      understand an entity  
\begynd
\pind that can be observed or conceived and described as a ``complete
      thing'' at no matter which given snapshot of time; 
\pind alternatively an entity is endurant if it is capable of
      \sfsl{enduring}, that is \sfsl{persist}, \sfsl{``hold out''}\pos{
      \cite[Vol.\,I, pg.\,656]{OED}}{}.
\afslut
Were we to ``freeze'' time 
\begynd
\pind we would still be able to observe the entire endurant \eod
\afslut
\afslut
}
\mnewfoil

\bookdefn{Perdurant}{%
\begynd
\pind By a \pdindextermi{perdurant} we shall understand an
      entity  
\begynd
\pind for which only a fragment exists\pos{}{\\} if we look at or
      touch them\pos{}{\\} at
      any given snapshot in time, that is, 
\pind were we to freeze time we would only see or
      touch \pos{}{\\} a fragment of the perdurant,
\pind alternatively
\begynd
\pind an entity is perdurant
\pind if it endures continuously, over time, persists, lasting\pos{
      \cite[Vol.\,II, pg.\,1552]{OED}}{} \eod 
\afslut
\afslut
\afslut
} 
\nbbbb{Endurants: Parts, Components and Materials}

\begynd
\pind  Endurants are either \bbcolor{\sfsl{discrete}} or
       \bbcolor{\sfsl{continuous}}.
\begynd
\pind  With discrete endurants we can choose to associate, or
       to not associate \sfsl{mereologies}\footnote{\LLLL\sf ---
         `mereology' will
         be explained next}.
\begynd
\pind  If we do we shall refer to them as \bbcolor{\sfsl{parts}},
\pind  else we shall call them \bbcolor{\sfsl{components}}.
\afslut
\pind With continuous endurants we do not associate mereologies.
\pind The continuous endurants we shall also refer to as \nyl
      \sfsl{(gaseous or liquid) \bbcolor{materials}.}
\afslut
\afslut
\begynd
\pind Parts are either
      \bbcolor{\sfsl{atomic}} or \bbcolor{\sfsl{composite}} and all parts have 
\begynd
\pind \sfsl{unique identifiers},
\pind \sfsl{mereology} and
\pind \sfsl{attributes}.
\afslut
\afslut
\dbeat{\pos{
\begynd
\pind Atomic parts \sfsl{may} \bbcolor{have} one or more \bbcolor{materials}
\begynd
\pind in which case we may observe these
      materials: \brcolor{obs\_materials}$(p)$
\pind which yields the informal and formal description:
\afslut
\afslut
\boks{Materials}{
\begynd
  \pind \textbf{Narrative:}
  \begynd
    \pind  ...
  \afslut
  \pind  \textbf{Formal:} 
  \begynd
    \pind \kw{type} 
    \begynd
      \pind $M_1, M_2, ..., M_n$
    \afslut
    \pind \kw{value}
    \begynd
      \pind \textsf{obs\_M$_i$: $P$ {\RIGHTARROW} $M_i$}
    \afslut
  \afslut repeated for all $n$ $M_i$s\,!
\afslut
}}{}}
\mnewfoil
\noindent
\begynd
\pind If the observed part, $p$:$P$, \bbcolor{is\_composite}
\begynd
\pind then we can observe the part sorts and values,
      $P_1, P_2, ..., P_m$ \nyl respectively $p_1, p_2, ..., p_m$ of $p$.
\pind ``Applying'' \brcolor{observe\_part\_sorts} to $p$ \nyl
      yields 
\begynd
\pind an informal (i.e., a \textbf{narrative}) and 
\pind a \textbf{formal} description:
\afslut
\afslut
\afslut
\mnewfoil
\boks{Composite Parts}{
\begynd
\pind \textbf{Narrative:}
\begynd
\pind ... 
\afslut
\pind \textbf{Formal:} 
\begynd
\pind \kw{type} 
\begynd
\pind $P_1, P_2, ..., P_m$,
\afslut 
\pind \kw{value}
\begynd
\pind \textsf{obs\_P$_i$: $P$ {\RIGHTARROW} $P_i$},
\afslut 
\afslut repeated for all $m$ part sorts $P_i$s\,!
\afslut
}

\mnewfoil

\nrslframebox{The Pragmatics}{%
\noindent
\begynd
\pind The \sfsl{pragmatics}\footnotemark\ of this ongoing example is this:
\begynd
\pind We are dealing with ordinary passenger aircraft.
\pind We are focusing on that tiny area of concern that \nyl focus on
passengers being informed of the progress of the flight, \nyl once in the air:
\begynd
\pind where is the aircraft: 
\begynd
\pind its current position somewhere above the earth;
\pind its current speed and direction 
\pind and possible acceleration (or deceleration); 
\pind We do not bother about what time it is -- etc.
\pind We abstract from \nyl the concrete presentation of this information.
\afslut
\afslut
\afslut
\afslut
}
\footnotetext{Pragmatics is here used in the sense outlined in
\cite[Chapter 7, Pages\,145--148]{TheSEBook2wo}.}

\mnewfoil

\nrslframebox{Parts}{
\begin{enumerate}\setei
\item \label{ip000} An \sfsl{aircraft} is composed from several parts
                    of which we focus on 
\begin{enumerate}
\item \label{ip001} a \sfsl{position} part,
\item \label{ip002}  a \sfsl{travel dynamics} part, and 
\item \label{ip003} a \sfsl{display} part. 
\end{enumerate}
\savei\end{enumerate}
\bp
\kw{type}\\
\ref{ip000}\ \ \ \ \ AC, PP, TD, DP\\
\kw{value}\\
\ref{ip001}\ \ \ obs\_PP: AC {\RIGHTARROW} PP\\
\ref{ip002}\ \ \ obs\_TD: AC {\RIGHTARROW} TD\\
\ref{ip003}\ \ \ obs\_DP: AC {\RIGHTARROW} DP 
\ep
}

\mnewfoil

\noindent
\begynd
\pind We have just summarised the analysis and description aspects of
endurants in \sfsl{extension} (their ``form'').
\pind We now summarise   the analysis and
description aspects of endurants in \sfsl{intension} (their ``contents'').
\pind There are three kinds of intensional \sfsl{qualities} associated with parts, two
with components, and one with materials.
\begynd
\pind Parts and components, by definition, have
\sfsl{unique identifiers}; 
\pind  parts have \sfsl{mereologies},
\pind  and all
{endurants} have \sfsl{attributes}.
\afslut
\afslut

\nbbbb{Internal Qualities}

\bbb{Unique Identifiers}

\begynd
\pind Unique identifiers are further
      undefined tokens that uniquely identify parts and
      components.
\pind The description language observer \brcolor{uid\_P}, when
      applied to parts $p$:$P$ yields the unique identifier, $\pi$:$\Pi$, of $p$.
\pind So the $\brcolor{observe\_part\_sorts}(p)$ invocation also
      yields the description text:
\afslut
\mnewfoil
\LLLL\HHHH
\boks{Unique Identifiers}{
\begynd
\pind ... [added to the narrative and]
\pind \kw{type}
\begynd
\pind $\Pi_1, \Pi_2, ..., \Pi_m$;
\afslut
\pind \kw{value}
\begynd
\pind \textsf{uid\_}$\Pi_i: P_i$\RIGHTARROW $\Pi_i$,    
\afslut repeated for all $m$ part sorts $P_i$s and added to the formalisation. 
\afslut
}

\mnewfoil
\nrslframebox{Unique Identifiers}{
\begin{enumerate}\setei
\item \label{ip010} position, travel dynamic and display parts have 
                    unique identifiers.
\savei\end{enumerate}
\bp
\kw{type}\\
\ref{ip010}\ \ \ PPI, TDI, DPI\\
\kw{value}\ \ \ \\
\ref{ip010}\ \ \ uid\_PP: PP {\RIGHTARROW} PPI\\
\ref{ip010}\ \ \ uid\_TD: TD {\RIGHTARROW} TDI\\
\ref{ip010}\ \ \ uid\_DP: DP {\RIGHTARROW} DPI\ \ 
\ep
}

\nbbb{Mereology}

\begynd
\pind \sfsl{Mereology is the study and knowledge of parts and
part relations.}
\begynd
\pind The mereology of a part is an expression over the
      unique identifiers of the (other) parts with which it is related,
\pind hence \brcolor{mereo\_P:}
      \textsf{$P${\RIGHTARROW}$\mathcal{E}(\Pi_j,...,\Pi_k)$}
      where $\mathcal{E}(\Pi_j,...,\Pi_k)$ is a type expression.
\pind So the $\brcolor{observe\_part\_sorts}(p)$ invocation also yields the
      description text:
      \afslut
      \mnewfoil
      
\boks{Mereology}{
\begynd
\pind ... [added to the narrative and]
\pind \kw{value}
\begynd
\pind \textsf{mereo\_}$P_i:
      P_i$\RIGHTARROW$\mathcal{E}_i(\Pi_{i_j},...,\Pi_{i_k})$  [added to the formalisation]
\afslut
\afslut
}
\afslut

\mnewfoil

\nrslframebox{Mereology}{
\noindent
\begynd
\pind We shall omit treatment of aircraft mereologies.
\afslut
\begin{enumerate}\setei
\item \label{mer10} The position part is related to the display part.
\item \label{mer20} The travel dynamics part is related to the display
                    part.
\item \label{mer30} The display part is related to both \nyl the position
  and the travel dynamics parts. 
\savei\end{enumerate}
\bp
\kw{value}\\
\ref{mer10}\ \ mereo\_PP: PP {\RIGHTARROW} DPI\\
\ref{mer20}\ \ mereo\_TD: TP {\RIGHTARROW} DPI\\
\ref{mer20}\ \ mereo\_DP: DP {\RIGHTARROW} PPI{\TIMES}TDI
\ep
}

\nbbb{Attributes}\label{Attributes}

\begynd
\pind Attributes are the remaining qualities of endurants.
\begynd
\pind The analysis prompt \brcolor{obs\_attributes} applied to an
      endurant yields a set of type names, $A_1, A_2, ..., A_t$, of
      attributes. 
\pind They imply the additional description text:
\afslut
\afslut
\mnewfoil
\boks{Attributes}{
\begynd
\pind \textbf{Narrative:}
\begynd
\pind ... 
\afslut
\pind \textbf{Formal:} 
\begynd
\pind \kw{type} 
\begynd
\pind $A_1, A_2, ..., A_t$
\afslut
\pind \kw{value}
\begynd
\pind \textsf{attr\_A$_i$: $E$ {\RIGHTARROW} $A_i$}\label{attr-A}
\afslut repeated for all $t$ attribute sorts $A_i$s\,!
\afslut
\afslut
}

\mnewfoil

\nrslframebox{Position Attributes}{
\begin{enumerate}\setei
\item \label{ip100} Position parts have longitude, latitude and
  altitude attributes.
\savei\end{enumerate}
\bp
\kw{type}\\
\ref{ip100}\ \ \ LO, LA, AL\\
\kw{value}\\
\ref{ip100}\ \ \ attr\_LO: PP {\RIGHTARROW} LO\\
\ref{ip100}\ \ \ attr\_LA: PP {\RIGHTARROW} LA\\
\ref{ip100}\ \ \ attr\_AL: PP {\RIGHTARROW} AL
\ep
\noindent
\begynd
\pind These quantities: longitude, latitude and altitude
\begynd
\pind are ``actual'' quantities, they mean what they express,
\pind they are not $r$ecordings or $d$isplays of these quantities;
\pind to express those we introduce separate types.
\afslut
\afslut
}

\mnewfoil

\nrslframebox{Travel Dynamics Attributes}{

\begin{enumerate}\setei
\item \label{ip110} Travel dynamics parts have velocity\footnotemark\ and
  acceleration\footnotemark.
\savei\end{enumerate}

\bp
\kw{type}\ \ \ \ \ \ \\
\ref{ip110}\ \ \ VEL, ACC\\
\kw{value}\\
\ref{ip110}\ \ \ attr\_VEL: TD {\RIGHTARROW} VEL\\
\ref{ip110}\ \ \ attr\_ACC: TD {\RIGHTARROW} ACC
\ep
\noindent
\begynd
\pind These quantities: velocity and acceleration,
\begynd
\pind are ``actual'' quantities, they mean what they express,
\pind they are not $r$ecordings or $d$isplays of these quantities;
\pind to express those we introduce separate types.
\afslut
\afslut
}
\addtocounter{footnote}{-1}
\footnotetext{\LLLL
  Velocity is a \sfsl{vector} of \sfsl{speed} and \sfsl{orientation}
  (i.e., \sfsl{direction})}
\stepcounter{footnote}
\footnotetext{\LLLL Acceleration is a vector of change of
  speed per time unit and orientation.}
\mnewfoil

\nrslframebox{Quantity Recordings}{
\begin{enumerate}\setei
\item \label{xian000} On one hand there are the actual location and
  dynamics quantities (i.e., values),
\item \label{xian010} on the other hand there are their recordings,
\item \label{xian020} and there are conversion functions from actual
                      to recorded values.
\savei\end{enumerate}
\bp
\kw{type}\\
\ref{xian000}\ \ \ LO, LA, AL, VEL, ACC\\
\ref{xian010}\ \ \ rLO, rLA, rAL, rVEL, rACC\\
\kw{value}\\
\ref{xian020}\ \ \ a2rLO: LO {\RIGHTARROW} rLO, a2rLA: LA {\RIGHTARROW} rLA, a2rAL: AL {\RIGHTARROW} rAL\\
\ref{xian020}\ \ \ a2rVEL: VEL {\RIGHTARROW} rVEL, a2rACC: ACC {\RIGHTARROW} rACC
\ep
\noindent
\begynd
\pind There are, of course, no functions that convert recordings to
actual values\,!
\afslut
}

\mnewfoil

\nrslframebox{Display Attributes}{\LLLL
\begin{enumerate}\setei
\item \label{ip120} Display parts have display-modified
  longitude, latitude and altitude, and velocity and
  acceleration attributes -- with functions
  that convert between these, recorded and displayed, attributes.
\savei\end{enumerate}
\begin{multicols}{2}\small
\bp
\kw{type}\\
\ref{ip120}\ \ \ dLO, dLA, dAL\\
\ref{ip120}\ \ \ dVEL, dACC\\
\kw{value}\ \ \ \ \ \\
\ref{ip120}\ \ \ attr\_dLO: DP {\RIGHTARROW} dLO\\
\ref{ip120}\ \ \ attr\_dLA: DP {\RIGHTARROW} dLA\\
\ref{ip120}\ \ \ attr\_dAL: DP {\RIGHTARROW} dAL\\
\ref{ip120}\ \ \ attr\_dVEL: DP {\RIGHTARROW} dVEL\\
\ref{ip120}\ \ \ attr\_dACC: DP {\RIGHTARROW} dACC\\
\ref{ip120}\ \ \ r2dLO,d2rLO: rLO $\leftrightarrow$ dLO\\
\ref{ip120}\ \ \ r2dLA,d2rLA: rLA $\leftrightarrow$ dLA\\
\ref{ip120}\ \ \ r2dAL,d2rAL: rAL $\leftrightarrow$ dAL\\
\ref{ip120}\ \ \ r2dVEL,d2rVEL: rVEL $\leftrightarrow$ dVEL\\
\ref{ip120}\ \ \ r2dACC,d2rACC: rACC $\leftrightarrow$ dACC\\
\kw{axiom}\\
\>\ {\ALL} rlo:rLO {\RDOT} d2rLO(r2dLO(rlo)){\EQ}rlo etcetera !
\ep
\end{multicols}
}

\nbbb{Attribute Categories}

\begynd
\pind Michael A.\ Jackson \cite{lexicon} categorizes and defines attributes  as either
\begynd
\pind \sfsl{static} or 
\pind \sfsl{dynamic}, 
\afslut
\pind with dynamic attributes being either
\begynd
\pind \sfsl{inert}, 
\pind \sfsl{reactive} or 
\pind \sfsl{active}. 
\afslut
\pind The latter are then either 
\begynd
\pind \sfsl{autonomous}, 
\pind \sfsl{biddable} or 
\pind \sfsl{programmable.}
\afslut
\pind This
      categorization has a strong bearing on how these (f.ex., part)
      attributes are dealt with when now interpreting parts as behaviours.
\afslut

\mnewfoil

\nrslframebox{Attribute Categories}{
\begin{enumerate}\setei
\item \label{ip200} Longitude, latitude, altitude, velocity and
  acceleration \nyl are all reactive attributes -- \nyl they
  change in response to the bidding of aircraft attributes \nyl that we
  have not covered\footnotemark.
\item \label{ip210} Their display modified forms are all programmable attributes.
\savei\end{enumerate}

\bp
\textbf{attribute categories}\\
\ref{ip200}\ \ \ \kw{reactive:}\ \ \ \ \ \ \ \ \ \ \ \ LO,LA,AL,VEL,ACC\\
\ref{ip210}\ \ \ \kw{programmable:} dLO,dLA,dAL,dVEL,dACC
\ep
}
\footnotetext{\LLLL -- for example: \sfsl{thrust, weight, lift, drag,
    rudder position}, and
  \sfsl{aileron position} -- plus dozens of other -- attributes}
\label{Attributes.n}

\nbbbb{Description Axioms and Proof Obligations}

\begynd
\pind In \cite{BjornerFAoC2015MDAAD} we show that the
      description prompts may result in axioms or proof obligations.
\begynd
\pind We refer to \cite{BjornerFAoC2015MDAAD} for details.
\pind Here we shall, but show one example of an axiom.
\afslut
\afslut

\mnewfoil

\pos{}{\vspace*{-5mm}}

\nrslframebox{An Axiom}{
\begin{enumerate}\setei
\item \label{ip300} The displayed attributes must at any time
                    be displays of the corresponding recorded position
                    and travel dynamics attributes.
\savei\end{enumerate}\LLLL
\bp
\kw{axiom} \\
\ref{ip300}\ \ \ {\ALWAYS} {\ALL} ac:AC {\RDOT} \\
\ref{ip300}\ \ \ \ \ \ \kw{let} (pp,td,di) {\EQ} (obs\_PP(ac),obs\_TD(ac),obs\_DP(ac)) \kw{in}\\
\ref{ip300}\ \ \ \ \ \ \kw{let} (lo,la,at) {\EQ} (attr\_LO(pp),attr\_LA(pp),attr\_AT(pp)),\\
\ref{ip300}\ \ \ \ \ \ \ \ \ \ \ (vel,acc,dir) {\EQ} (attr\_VEL(td),obs\_ACC(td)),\\
\ref{ip300}\ \ \ \ \ \ \ \ \ \ \ (dlo,dla,dat) {\EQ} (attr\_dLO(di),attr\_dLA(di),attr\_dAT(di)),\\
\ref{ip300}\ \ \ \ \ \ \ \ \ \ \ (dvel,dacc) {\EQ} (attr\_dVEL(di),obs\_dACC(di)) \kw{in}\\
\ref{ip300}\ \ \ \ \ \ (dlo,dla,dat) {\EQ} (r2dLO(a2rLO(lo)),r2dLA(a2rLA(la)),r2dAL(a2rAL(at)))\\
\ref{ip300}\ \ \ {\WEDGE} (dvel,dacc) {\EQ} (r2dVEL(a2rVEL(vel)),r2dACC(a2rACC(acc)))\\
\ref{ip300}\ \ \ \ \ \ \kw{end} \kw{end}\ \ \ \ \ \ \ 
\ep
}

\nbbbb{From Manifest Parts (Endurants) to Domain Behaviours (Perdurants)}

\begynd
\pind \cite{BjornerFAoC2015MDAAD} then presents a \sfsl{compiler}
      which to manifest \sfsl{parts} associate
      \sfsl{behaviours}. 
\pind These are then specified as \texttt{CSP} \citecsp\
      \sfsl{processes}. 
\begynd
\pind We choose \texttt{CSP} \cite{Hoa78a,Hoare85,CARH:Electronic,Roscoe97,Schneider99} for
      the following reasons:  
\begynd
\pind it is a well-established formalism for expressing
      the behaviour of cooperating sequential processes;
\pind it has withstood the test of time: first articles
      appeared in 1978 and research and industrial use is still at a
      high; 
\pind it has a well-founded theory and a
      \sfsl{failures--divergence-refinement} proof system with proof
      rules \cite{RoscoeFDR94a,FDR2:2004};
\pind we have shown, in \cite[To Every Manifest Domain a
      \texttt{CSP} Expression -- A R{\^o}le for Mereology in Computer
      Science]{BjornerMereologyCSP2017}, how to relate a world of
      space-based endurants with a world of  time-based perdurants;
\pind in \cite[A Philosophy of Domain Science \&
  Engineering --An Interpretation of Kai 
  S{\o}rlander's Philosophy]{2018:Bjorner:philo}.
\afslut
\mnewfoil
\pind The latter two reasons, to us, are rather convincing:
\begynd
\pind The concept of \sfsl{mereology} is well-established,
      \nyl both in philosophy and in logic.
\pind To ``connect'' the concepts of mereology with ontological
      concepts of describable domains, by some \texttt{neo-Kantian}
      \sfsl{transcendental deduction} is quite surprising -- and pleasing. 
\afslut
\afslut
\afslut
      
\nbbb{The Idea --- by means of an example}

\begynd
\pind The term \sfsl{aircraft} can have the following ``meanings'':
\begynd
\pind the \sfsl{aircraft}, as an \sfsl{endurant}, \nyl parked at the airport
      gate, \nyl i.e., as a \sfsl{composite part;}
\pind the  \sfsl{aircraft}, as a \sfsl{perdurant}, \nyl as it flies
      through the skies, \nyl i.e., as a \sfsl{behaviour;} and 
\pind the \sfsl{aircraft}, as an \sfsl{attribute}, \nyl of an airline timetable.
\afslut
\afslut

\mnewfoil

\nrslframebox{An Informal Story}{%
\noindent\begynd
\pind An aircraft has the following  behaviours:
\begynd
\pind the \sfsl{position} behaviour; \nyl it \underline{observes} 
      the aircraft location attributes: \nyl
      \sfsl{longitude, latitude} and \sfsl{altitude}, \nyl
      \underline{record} and \underline{communicate} these, as a
      triple, to the \sfsl{display}  behaviour; 
\pind the \sfsl{travel dynamics}  behaviour; \nyl it
      \underline{observes} the aircraft travel dynamics attributes
      \nyl \sfsl{velocity} and \sfsl{acceleration},
      \nyl \underline{record} 
      and \underline{communicate} these, as a triple, to the
      \sfsl{display}  behaviour; and 
\pind the \sfsl{display} behaviour \underline{receives} two doublets of attribute 
      value \underline{recordings} \nyl from respective
      \sfsl{position} and  \sfsl{travel dynamics}  behaviours \nyl
      and \underline{display} these recorded attribute values: \nyl
      \sfsl{longitude, latitude, altitude, velocity} and 
      \sfsl{acceleration} in some form.
\afslut
\afslut
\mnewfoil
\pos{}{\vspace*{-6mm}}\LLLL
      \begin{center}
        \epsfig{file=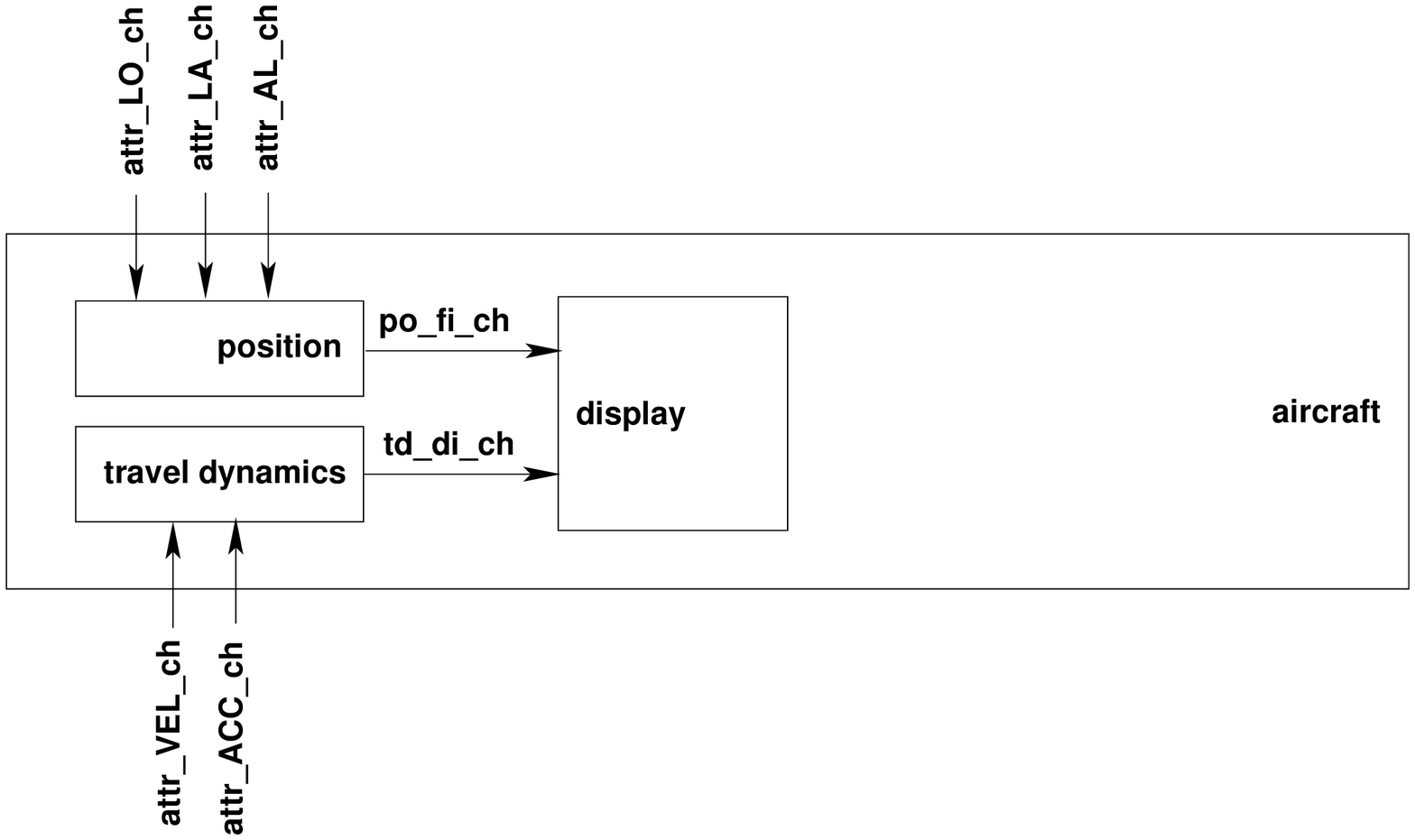,width=\pos{12}{16}cm}
      \end{center}
\noindent\LLLL
\begynd
\pind The six \underline{actual} \sfsl{position} and \sfsl{travel
      dynamics} attribute values 
      \nyl \sfsl{longitude, latitude, altitude, velocity} and
      \sfsl{acceleration}
\begynd
\pind are \underline{recorded}, by appropriate instruments.
\pind In the above figure this is indicated by \kw{in}put channels
      \nyl \textsf{attr\_LO\_ch, attr\_LA\_ch, attr\_AL\_ch,
        attr\_VEL\_ch} and \textsf{attr\_ACC\_ch}.
\afslut
\afslut
}

\dbeat{\pos{\nrslframebox{An Informal Story}{\LLLL\input{story1}\mnewfoil\input{story2}}}%
     {\nrslframebox{An Informal Story I/II}{\LLLL\input{story1}}\mnewfoil\LLLL
      \unrslframebox{An Informal Story II/II}{\LLLL\input{story2}}}}

\nbbb{Channels and Communication}

\noindent
\begynd
\pind Behaviours sometimes synchronise and usually communicate.
\pind We use the \texttt{CSP} \citecsp\
      notation (adopted by \texttt{RSL}) to  model behaviour
      communication. 
\pind Communication is abstracted as the sending,
\begynd
\pind \textsf{ch\,!\,m},  \pos{}{} and receipt,
\pind \textsf{ch\,?},  \pos{}{} of messages,
\pind \textsf{m:M}, \pos{}{}  over channels,
\pind \textsf{ch}. 
\afslut
\afslut
\bp
\>\>\kw{type} M\\
\>\>\kw{channel} ch:M
\ep

\mnewfoil\LLLL\HHHH

\nrslframebox{Channels}{
\pos{\noindent
\begynd
\pind For this example we focus only on communications from \nyl
      the \sfsl{position} and \sfsl{travel dynamics} behaviours \nyl 
      to the \sfsl{display} behaviour.
\afslut}{}
\begin{enumerate}\setei
\item \label{xian-chan-010} The messages sent from the \sfsl{position}
                            behaviour \nyl to the \sfsl{display}
                            behaviour \nyl are triplets of \underline{recorded}
                            longitude, latitude and altitude values.
\item \label{xian-chan-020} The messages sent from the \sfsl{travel dynamics}
                            behaviour \nyl to the \sfsl{display}
                            behaviour \nyl are droplets of
                            \underline{recorded}
                            velocity and acceleration values.
\item \label{xian-chan-030} There is a channel, {\sf po\_di\_ch}, that
                            allows communication of messages \nyl from the
                            position behaviour to the display behaviour.
\item \label{xian-chan-040} There is a channel, {\sf td\_di\_ch}, that
                            allows communication of messages \nyl from the
                            travel dynamics behaviour to the display behaviour.
\item \label{xian-chan-041} For each of the  reactive attributes there
                            is a corresponding channel. 
\savei\end{enumerate}
\mnewfoil
\bp
\kw{type}\\
\ref{xian-chan-010}\ \ \ PM {\EQ} rLO {\TIMES} rLA {\TIMES} rAL\\
\ref{xian-chan-020}\ \ \ TDM {\EQ} rVEL {\TIMES} rACC\\
\kw{channel}\ \ \\
\ref{xian-chan-030}\ \ \ po\_di\_ch:PM\\
\ref{xian-chan-040}\ \ \ td\_di\_ch:TDM \\
\ref{xian-chan-041}\ \ \ attr\_LO\_ch:LO, attr\_LA\_ch:LA, attr\_AL\_ch:AL\\
\ref{xian-chan-041}\ \ \ attr\_VEL\_ch:VEL, attr\_ACC\_ch:ACC
\ep
}
\dbeat{\pos{\nrslframebox{Channels}{\input{ch1}\mnewfoil\input{ch2}}}%
     {\nrslframebox{Channels I/II}{\input{ch1}}\mnewfoil
      \unrslframebox{Channels II/II}{\input{ch2}}}}

\nbbb{Behaviour Signatures}\LLLL\HHHH\label{DBSaD}\label{ss.Signatures.4} 

\begynd
\pind We shall only cover behaviour signatures
      when expressed in \texttt{RSL/CSP} \citersl.
\pind The behaviour functions are now called processes.
\pind That a behaviour function is a never-ending function, i.e., a
      process,\pos{}{\\} is ``revealed'' in the function signature by
      the ``trailing'' {\textbf{{Unit}}}:
\bp
\>\>\>behaviour: {\DOTDOTDOT} {\RIGHTARROW} {\DOTDOTDOT} \kw{Unit}
\ep
\noindent
\pind That a process takes no argument is ''revealed'' by a
``leading''  \textsf{Unit}:
\bp
\>\>\>behaviour: \kw{Unit} {\RIGHTARROW} {\DOTDOTDOT}
\ep
\noindent

\pind That a process accepts channel, viz.: \textsf{ch},
      inputs, including accesses an external attribute \textsf{A}, is
      ``revealed'' in the function signature as follows:
\bp
\>\>\>behaviour: {\DOTDOTDOT} {\RIGHTARROW} \kw{in} ch {\DOTDOTDOT}\ \ , resp.\  \kw{in} attr\_A\_ch
\ep
\mnewfoil
\noindent 
\pind That a process offers channel, viz.: \textsf{ch}, outputs  is
      ``revealed'' in the function signature as follows:
\bp
\>\>\>behaviour: {\DOTDOTDOT} {\RIGHTARROW} \kw{out} ch {\DOTDOTDOT} 
\ep
\noindent
\pind That a process accepts other arguments   is
      ``revealed'' in the function signature as follows:
 
\bp
\>\>\>behaviour: ARG {\RIGHTARROW} {\DOTDOTDOT} 
\ep
\noindent
\pind where \textsf{ARG} can be any type expression:
\bp
\>\>\>T, T{\RIGHTARROW}T, T{\RIGHTARROW}T{\RIGHTARROW}T, etcetera\ \ \ 
\ep
\noindent where \textsf{T} is any type expression.
\faoce
\afslut

\nbbb{Translation of Part Qualities}

\begynd
\pind  Part qualities, that is: \sfsl{unique identifiers, mereologies} and
       \sfsl{attributes}, are translated into behaviour arguments -- of one
       kind or another, i.e., (...).
\begynd
\pind Typically we can choose to \sfsl{index} behaviour
      names, $b$ by the \sfsl{unique identifier}, $id$, of the part based on
      which they were translated, i.e., $b_{id}$.
\pind \sfsl{Mereology values} are usually static, and can, as thus, be
      treated like we treat static attributes (see next), or can be
      set by their behaviour, and are then treated like we treat
      programmable attributes (see next), i.e., (...).
\mnewfoil
\pind \sfsl{Static attributes} become behaviour definition (body) constant values.
\pind \sfsl{Inert, reactive} and \sfsl{autonomous attributes} become
      references to channels, say \sfsl{ch\_dyn}, such that when an 
      inert, reactive and autonomous attribute value is required it is
      expressed as \sfsl{ch\_dyn\,?}.
\pind \sfsl{Programmable} and \sfsl{biddable attributes} become  arguments
      which are passed on to the tail-recursive invocations of the
      behaviour, and possibly updated as specified [with]in the body of the
      definition of the behaviour, i.e., (...).
\afslut
\afslut

\nbbb{Part Behaviour Signatures}

\begynd
\pind We can, without loss of 
      generality, associate with each part a behaviour;  
\begynd 
\pind parts which share attributes 
\pind (and are therefore referred to in some parts' mereology),
\pind can communicate (their ``sharing'') via channels.
\afslut
\mnewfoil
\pind A behaviour signature is therefore:\HHHH

\bp
\>\>\>beh$_{\pi:\Pi}$: me:MT{\TIMES}sa:SA{\RIGHTARROW}ca:CA{\RIGHTARROW}\kw{in} $ichns$(ea:EA) \kw{in},\kw{out} $iochs$(me) \kw{Unit}
\ep

\noindent\HHHH where
\begynd
\pind (i)   $\pi$:$\Pi$ is the unique identifier of part \textsf{p}, i.e.,
            \textsf{$\pi$={\uidmo}P(p)}, 
\pind (ii)  \textsf{me:ME} is the mereology of part \textsf{p}, \textsf{me} =
            \textsf{\mereomo{P}(p)}, 
\pind (iii) \textsf{sa:SA} lists the static attribute values of the part, 
\pind (iv)  \textsf{ca:CA} lists the biddable and programmable
            attribute values of the part, 
\pind (v)  \bnovchg\textsf{$ichns$(ea:EA)} refer to the  external attribute
            $i$nput $ch$a$n$nel$s$, and where
\pind (vi) \textsf{$iochs$(me)} are the input/output channels serving
           the attributes shared between the part \textsf{p} and the
           parts designated in its mereology \textsf{me}.
\afslut
\afslut
\mnewfoil

\begynd
\pind We focus, for a little while, on the expression of 
\pos{}{\begin{multicols}{3}}
\begynd
\pind \textsf{sa:SA},
\pind \textsf{ea:EA} and
\pind \textsf{ca:CA},
\afslut
\pos{}{\end{multicols}}
\pind that is, on the concrete types of \textsf{SA, EA} and
      \textsf{CA}.
\begynd
\pind \textsf{sa:SA} lists the static value types,
      (\(svT_1, ..., svT_s \)),\pos{}{\\} where $s$ is the number of $s$tatic
      attributes of parts p:P.
\pind\textsf{ea:EA} lists the external
      attribute value channels of parts $p$:$P$ in the behaviour
      signature and as input channels, \textsl{ichns}, see 9 lines above.
\pind \textsf{ca:CA} lists the 
      controllable \enovchg value 
      expression types of parts $p$:$P$.      
\begynd
\pind A \pdindextermiii{controllable}{attribute}{value expression} is an
      expression involving one or more 
      attribute value expressions of the type of the biddable or
      programmable attribute\ \eod\ 
\afslut
\afslut
\afslut

\mnewfoil\LLLL\HHHH

\nrslframebox{Part Behaviour Signatures, I/II}{\pos{\noindent
\begynd
\pind We omit the signature of the aircraft behaviour.
\afslut
}{}
\begin{enumerate}\setei
\item \label{pbs0} The signature of the \sfsl{position} behaviour
  lists its \nyl unique identifier, mereology, no static and no controllable attributes,
  \nyl but its three reactive attributes (as input channels) \nyl and its (output)
  channel to the \sfsl{display} behaviour.
\item \label{pbs1} The signature of the \sfsl{travel dynamics} behaviour
  lists its \nyl unique identifier, mereology, no static and no controllable attributes,
  \nyl but its three reactive attributes (as input channels) \nyl and its (output)
  channel to the \sfsl{display} behaviour.. 
\item \label{pbs2} The signature of the \sfsl{display} behaviour
  lists its \nyl unique identifier, its mereology, no static attribute, but
  the programmable display attributes, assembled in a pair of a
  triplet and doublets,  and its two input channels
  from the \sfsl{position}, respectively the \sfsl{travel dynamics} behaviours.
\savei\end{enumerate}
}

\mnewfoil

\nrslframebox{Part Behaviour Signatures, I/II}{

\bp
\kw{type}\\
\ref{pbs2}\ \ \ DA {\EQ} (dLA{\TIMES}dLO{\TIMES}dAL){\TIMES}(dVEL{\TIMES}dACC)\\
\kw{value}\\
\ref{pbs0}\ \ \ position: PI {\TIMES} DPI {\RIGHTARROW} \\
\ref{pbs0}\ \ \ \ \ \ \ \ \kw{in} attr\_LO\_ch,attr\_LA\_ch,attr\_AL\_ch, \kw{out} po\_di\_ch\ \ \kw{Unit}\\
\ref{pbs1}\ \ \ travel\_dynamics: TDI {\TIMES} DPI {\RIGHTARROW}\ \ \\
\ref{pbs1}\ \ \ \ \ \ \ \ \kw{in} attr\_VEL\_ch,attr\_ACC\_ch,attr\_DIR\_ch, \kw{out} td\_di\_ch\ \ \kw{Unit}\\
\ref{pbs2}\ \ \ display: DI {\TIMES} (PPI{\TIMES}TDI) {\RIGHTARROW} DA {\RIGHTARROW} \kw{in} po\_di\_ch, td\_di\_ch\ \ \kw{Unit} 
\ep

}

\nbbb{Behaviour Compilations}
\bb{Composite Behaviours}

\begynd
\pind Let \textsf{P} be a composite sort defined in terms of sub-sorts
      \textsf{P$_1$, P$_2$, \ldots, P$_n$}.
\begynd
\pind The process definition compiled from \textsf{p:P}, is composed from 
\begynd
\pind a process description, $\mathcal{M}{{cP}}_{{\uidmo}P(p)}$,
      relying on and handling the 
      unique identifier, mereology and attributes of part $p$ 
\pind operating in parallel with processes $p_1, p_2, \ldots, p_n$ where 
\begynd
\pind $p_1$ is compiled from \textsf{p$_1$:P$_1$}, 
\pind  $p_2$ is compiled from \textsf{p$_2$:P$_2$}, 
\pind ..., and 
\pind $p_n$ is compiled from \textsf{p$_n$:P$_n$}.
\afslut
\afslut
\afslut
\pind The domain description ``compilation'' schematic below
      ``formalises'' the above. 
\afslut
\mnewfoil\LLLL\HHHH
\nramme{\kw{Transcendental Schema: Abstract
    \bbcolor{\texttt{is\_composite(p)}}}\label{Process Schema
    I}}\label{psI} 
\bp
\>\>\kw{value}\\
\>\>\>\>compile\_process: P {\RIGHTARROW} \textsf{RSL}-\kw{Text}\\
\>\>\>\>compile\_process(p) {\IS} \\
\>\>\>\>\>\>\>\ $\mathcal{M}{{{P}}}_{{\uidmo}P(p)}$(\mereomo{P}(p),$\mathcal{S_A}$(p))($\mathcal{C_A}$(p))\\
\>\>\>\>\>\>\>{\PARL} compile\_process({\obsmo}P$_1$(p))\\
\>\>\>\>\>\>\>{\PARL} compile\_process({\obsmo}P$_2$(p))\\
\>\>\>\>\>\>\>{\PARL} {\DOTDOTDOT}\\
\>\>\>\>\>\>\>{\PARL} compile\_process({\obsmo}P$_n$(p))
\ep
\noindent
\begynd
\pind The text macros: $\mathcal{S_A}$ and
      $\mathcal{C_A}$ were informally explained above.
\pind Part sorts \textsf{P$_1$, P$_2$, ..., P$_n$} are obtained \nyl from the
      \texttt{observe\_part\_sorts} prompt. 
\afslut
\nemmar
\mnewfoil

\nrslframebox{Aircraft Behaviour, I/II}{
\begin{enumerate}\setei
\item \label{ip310} Compiling a composite aircraft part \nyl results  in
  the parallel composition
  \begin{enumerate}
\item \label{ip330} the compilation of the atomic position part,
\item \label{ip340} the compilation of the atomic travel dynamics  part, and
\item \label{ip350} the compilation of the atomic display part.
\end{enumerate}
We omit compiling the aircraft core behaviour.
\item \label{ip360} Compilation of atomic parts entail no further compilations.
\savei\end{enumerate}
}
\mnewfoil

\unrslframebox{Aircraft Behaviour, II/II}{
\noindent
\bp
\kw{value}\\
\ref{ip310}\ \ \ compile(ac) {\IS}\\
\ref{ip330}\ \ \ \ \ \ \ compile(obs\_PP(p))\\
\ref{ip340}\ \ \ \ {\PARL} compile(obs\_TD(p))\\
\ref{ip350}\ \ \ \ {\PARL} compile(obs\_DI(p))
\ep
}

\nbb{Atomic Behaviours}

\HHHH
\nramme{\kw{Transcendental Schema:
    \bbcolor{\texttt{is\_atomic(p)}}}\label{Process Schema III}}\label{psIII} 
\bp
\>\>\kw{value}\\
\>\>\>\>compile\_process: P {\RIGHTARROW} \textsf{RSL}-\kw{Text}\\
\>\>\>\>compile\_process(p) {\IS} \\
\>\>\>\>\>\>\>$\mathcal{M}{{P}}_{{\uidmo}P(p)}$(\mereomo{P}(p),$\mathcal{S_A}$(p))($\mathcal{C_A}$(p))
\ep
\nemmar

\mnewfoil

\nrslframebox{Atomic Behaviours}{

\begin{enumerate}\setei
\item \label{ip35x} We initialise the display behaviour with a further undefined value.
\savei\end{enumerate}
\bp
\kw{value}\ \ \\
\ref{ip330}\ \ \ \ compile(obs\_PP(p)){\IS} \\
\ref{ip330}\ \ \ \ \ \ \ \ \ position(uid\_PP(p),mereo\_PP(p))\\
\ref{ip340}\ \ \ \ compile(obs\_TD(p)) {\IS} \\
\ref{ip340}\ \ \ \ \ \ \ \ \ travel\_dynamics(uid\_TD(p),mereo\_TD(p))\\
\ref{ip35x}\ \ \ \ \ init\_DA:DA {\EQ} {\DOTDOTDOT}\\
\ref{ip350}\ \ \ \ compile(obs\_DI(p)) {\IS} \\
\ref{ip350}\ \ \ \ \ \ \ \ \ display(.uid\_DI(p),mereo\_DI(p))(init\_DA)
\ep
\noindent
\begynd
\pind In the above we have already subsumed the \nyl \sfsl{atomic behaviour
  definitions}, \nyl see next, and directly inserted the $\mathcal{F}$ definitions.
\afslut
}

\nbbb{Atomic Behaviour Definitions}

\nramme{\kw{Transcendental Schema IV: Atomic Core Processes}\label{Process Schema}}\label{psIV}
\noindent             
%
\bp
\kw{value}\\
\>\ $\mathcal{M}{{P}}_{\pi:\Pi}$: me:MT{\TIMES}sa:SA {\RIGHTARROW} ca:CA {\RIGHTARROW} \\
\>\>\>\>\kw{in} $ichns$(ea:EA) \kw{in},\kw{out} $iochs$(me)\ \ \kw{Unit}\\
\>\ $\mathcal{M}{{P}}_{\pi:\Pi}$(me,sa)(ca) {\IS} \\
\>\>\>\>\kw{let} (me{\PRIM},ca{\PRIM}) {\EQ} $\mathcal{F}_{\pi:\Pi}$(me,sa)(ca) \kw{in} \\
\>\>\>\>$\mathcal{M}{P}_{\pi:\Pi}$(me{\PRIM},sa)(ca{\PRIM}) \kw{end}\\
\\
\>\ $\mathcal{F}_{\pi:\Pi}$: me:MT{\TIMES}sa:SA {\RIGHTARROW} CA {\RIGHTARROW} \\
\>\>\>\>\kw{in} $ichns$(ea:EA) \kw{in},\kw{out} $iochs$(me) {\RIGHTARROW} MT{\TIMES}CA
\ep
\nemmar

\mnewfoil

\nrslframebox{Position Behaviour Definition}{
\begin{enumerate}\setei
\item \label{pbd1} The \sfsl{position} behaviour offers to receive \nyl
                    the \sfsl{longitude}, \sfsl{latitude} and the
                    \sfsl{altitude} attribute values
\item \label{pbd2} and to offer them to the \sfsl{display} behaviour, 
\item \label{pbd3} whereupon it resumes being the \sfsl{position} behaviour.
\savei\end{enumerate}
\bp
\kw{value}\\
\ref{pbs0}\ \ \ position(p$\pi$,d$\pi$) {\IS}\\
\ref{pbd1}\ \ \ \ \ \ \ \kw{let} (lo,la,al) {\EQ} (attr\_LO\_ch?,attr\_LA\_ch?,attr\_AL\_ch?) \kw{in}\\
\ref{pbd2}\ \ \ \ \ \ \ po\_di\_ch ! (a2rLO(lo),a2rLA(la),a2rAL(al)) ;\\
\ref{pbd3}\ \ \ \ \ \ \ position(p$\pi$,d$\pi$) \kw{end}
\ep
}

\mnewfoil

\nrslframebox{Travel Dynamics Behaviour Definition}{
\begin{enumerate}\setei
\item \label{td1} The \sfsl{travel\_dynamics} behaviour offers to
                    receive \nyl
                    the recorded \sfsl{velocity} and the
                    \sfsl{acceleration} attribute values
\item \label{td2} and to offer these to the \sfsl{display} behaviour, 
\item \label{td3} whereupon it resumes being the \sfsl{travel\_dynamics} behaviour.
\savei\end{enumerate}\LLLL\HHHH
\bp
\kw{value}\\
\ref{pbs1}\ \ \ travel\_dynamics(td$\pi$,d$\pi$) {\IS}\\
\ref{td1}\ \ \ \ \ \ \ \kw{let} (vel,acc){\EQ}(attr\_VEL\_ch?,attr\_ACC\_ch?) \kw{in}\\
\ref{td2}\ \ \ \ \ \ \ td\_di\_ch ! (a2rVEL(vel),a2rACC(acc)) ;\\
\ref{td3}\ \ \ \ \ \ \ travel\_dynamics(td$\pi$,d$\pi$) \kw{end}
\ep
\HHHH
}

\mnewfoil\LLLL

\nrslframebox{Display Behaviour Definition}{
\begin{enumerate}\setei
\item \label{dib1} The \sfsl{display} behaviour offers \nyl to receive the
                   reactive attribute doublets \nyl from the \sfsl{position}
                   and the \sfsl{travel\_dynamics} behaviours while
\item \label{dib2} resuming to be that behaviour albeit now with these \nyl
                   as their updated display.
\item \label{dib3} The \textbf{conv}ersion functions are extensions
  of the ones introduced earlier.
\savei\end{enumerate}
\bp
\kw{value}\\
\ref{pbs2}\ \ \ display(d$\pi$,(d$\pi$,td$\pi$))(d\_pos,d\_tdy) {\IS}\\
\ref{dib1}\ \ \ \ \ \ \ \kw{let} (pos\_d$'$,tdy\_d$'$) {\EQ} (po\_di\_ch?,td\_di\_ch?) \kw{in}\\
\ref{dib2}\ \ \ \ \ \ \ display(d$\pi$,(d$\pi$,td$\pi$))(conv(pos\_d$'$),conv(c\_tdy\_d$'$)) \kw{end}\\
\kw{type}\\
\ref{dib3}\ \ \ dMPD {\EQ} dLO {\TIMES} dLA {\TIMES} dAL\\
\ref{dib3}\ \ \ dMTD {\EQ} dVEL {\TIMES} dACC\\
\kw{value}\\
\ref{dib3}\ \ \ conv: MPD {\RIGHTARROW} dMPD\\
\ref{dib3}\ \ \ conv(rlo,rla,ral) {\IS} (r2dLO(rlo),r2dLA(rla),r2dAL(ral))\\
\ref{dib3}\ \ \ conv: MTD {\RIGHTARROW} dMTD\\
\ref{dib3}\ \ \ conv(rvel,racc) {\IS} (r2dVEL(rvel),r2dACC(racc))
\ep
}

\nbbbb{A Proof Obligation}

\begynd
\pind We refer, again, to \cite{BjornerFAoC2015MDAAD} for more on proof obligations.
\afslut

\nrslframebox{A Proof Obligation}{%
\begynd
\pind The perdurant descriptions of Items\,\ref{xian-chan-010}--\ref{dib3} 
\pind is a model of the axiom expressed in Item\,\ref{ip300}.
\afslut}

\nbbbbb{Calculations in Classical Domains: Some Simple Observations}

\begynd
\pind This section \pos{}{of the talk} covers three loosely related
      topics:
\begynd
\pind \pos{Sect.\,\ref{soosav} muses}{First we muse} over properties
      of some attribute values.
\pind Then\pos{, Sect.\,\ref{si}}{} we recall some facts about types,
      scales and values of measurable units in physics.
\pind The previous leads us\pos{, in Sect.\,\ref{attrs:tsv}}{} to consider
      further detailing the concept of attributes such as we have covered
      it in Sect.\,\pos{\ref{Attributes}, Pages}{2.2.3,
      Slides\,\pageref{Attributes}--\pageref{Attributes.n},} and in 
      \cite{BjornerFAoC2015MDAAD}. 
\afslut
\pind The reason for covering these topics is that
\begynd
\pind most attribute values are represented in ``final'' programs \nyl
      as numbers of one kind or another 
\pind and that type checking in most software \nyl is with respect to
      these numbers. 
\afslut
\afslut

\nbbbb{Some Observations on Some Attribute Values}\label{soosav}

Let us, seemingly randomly, examine some simple, \nyl e.g.,
      arithmetic, operations in classical domains.
\begynd
\pind By \sfsl{time} is often meant absolute time.
\pind So a time could be \sfsl{\todaytime.}
\pind One can not add two times.\label{soosav-time}
\pind One can speak of a time being earlier, or before another time.
\pind \sfsl{October 23, 2017: 10:01 am} is earlier, $\leq$, \nyl than \sfsl{\todaytime.}
\pind One can speak of the time interval between \LLLL\HHHH
\begynd
\pind \sfsl{October 23, 2016: 8:01 am} and  \sfsl{October 24, 2017: 10:05 am} 
\pind being \sfsl{1 year, 1 day, 2 hours and 4 minutes,} that is:
      \sfsl{October 24, 2017: 10:05 am $\ominus$ October 23, 2016:
      8:01 am  = 1 year, 1 day, 2 hours and 4 minutes}
\afslut
\afslut
\mnewfoil
\noindent
\begynd
\pind One can add a \sfsl{time interval} to a \sfsl{time} and obtain a \sfsl{time}.
\pind One can multiply a \sfsl{time interval} with a \sfsl{real}\footnote{\LLLL
  The time interval could, e.g., be converted into seconds, then the
  integer number standing for seconds can be multiplied by $r$ and the
  result be converted ``back'' into years, days, hours, minutes and
  seconds --- whatever it takes\,!}
\pind We can formalize the above:\LLLL\LLll
\afslut

\pos{\begin{multicols}{2}}{}
\bp
\kw{type}\\
\>\ $\mathbb{T}$ {\EQ} Month{\TIMES}Day{\TIMES}Year{\TIMES}Hour{\TIMES}Minute{\TIMES}Sec{\DOTDOTDOT}\\
\>\ $\mathbb{TI}$ {\EQ} Days{\TIMES}Hours{\TIMES}Minutes{\TIMES}Seconds{\TIMES}{\DOTDOTDOT}\\
\>\ Month {\EQ} {\LBRACE}{\BAR}1,2,3,4,5,6,7,8,9,10,11,12{\BAR}{\RBRACE}\\
\>\ Day {\EQ} {\LBRACE}{\BAR}1,2,3,4,{\DOTDOTDOT},28,29,30,31{\BAR}{\RBRACE}\\
\>\ Hour,Hours {\EQ} {\LBRACE}{\BAR}0,1,2,3,{\DOTDOTDOT},21,22,23{\BAR}{\RBRACE}\\
\>\ Minute,Minutes {\EQ} {\LBRACE}{\BAR}0,1,2,3,{\DOTDOTDOT}.,56,57,58,59{\BAR}{\RBRACE}\\
\>\ Second,Seconds {\EQ} {\LBRACE}{\BAR}0,1,2,3,{\DOTDOTDOT}.,56,57,58,59{\BAR}{\RBRACE}\\
\>\ {\DOTDOTDOT}\\
\>\ Days {\EQ} \kw{Nat}
\ep
\pos{}{\mnewfoil\LLll\HHHH}
\pos{
\bp
\kw{value}\\
\>\ {\LT},{\LEQ},{\EQ},{\GEQ},{\GT}:\ \ $\mathbb{T}$ {\TIMES} $\mathbb{T}$ {\RIGHTARROW} \kw{Boole}\\
\>\ {\MINUS}: $\mathbb{T}$ {\TIMES} $\mathbb{T}$ {\RIGHTARROW} $\mathbb{TI}$ \kw{pre} t{\MINUS}t{\PRIM}: t{\PRIM}{\LEQ}t\\
\>\ {\LT},{\LEQ},{\EQ},{\GEQ},{\GT}:\ \ $\mathbb{TI}$ {\TIMES} $\mathbb{TI}$ {\RIGHTARROW} \kw{Bool}\\
\>\ {\MINUS},{\PLUS}: $\mathbb{TI}$ {\TIMES} $\mathbb{TI}$ {\RIGHTARROW} $\mathbb{TI}$\\
\>\ {\AST}:\ \ $\mathbb{TI}$ {\TIMES} \kw{Real} {\RIGHTARROW} $\mathbb{TI}$\\
\>\ /:\ \ $\mathbb{TI}$ {\TIMES} $\mathbb{TI}$ {\RIGHTARROW} \kw{Real}
\ep
}
{
\bp
\>\>\>\>\>\>\ \kw{value}\\
\>\>\>\>\>\>\>\>{\LT},{\LEQ},{\EQ},{\GEQ},{\GT}:\ \ $\mathbb{T}$ {\TIMES} $\mathbb{T}$ {\RIGHTARROW} \kw{Bool}\\
\>\>\>\>\>\>\>\>{\MINUS}: $\mathbb{T}$ {\TIMES} $\mathbb{T}$ {\RIGHTARROW} $\mathbb{TI}$ \kw{pre} t{\MINUS}t{\PRIM}: t{\PRIM}{\LEQ}t\\
\>\>\>\>\>\>\>\>{\LT},{\LEQ},{\EQ},{\GEQ},{\GT}:\ \ $\mathbb{TI}$ {\TIMES} $\mathbb{TI}$ {\RIGHTARROW} \kw{Bool}\\
\>\>\>\>\>\>\>\>{\MINUS},{\PLUS}: $\mathbb{TI}$ {\TIMES} $\mathbb{TI}$ {\RIGHTARROW} $\mathbb{TI}$\\
\>\>\>\>\>\>\>\>{\AST}:\ \ $\mathbb{TI}$ {\TIMES} \kw{Real} {\RIGHTARROW} $\mathbb{TI}$\\
\>\>\>\>\>\>\>\>/:\ \ $\mathbb{TI}$ {\TIMES} $\mathbb{TI}$ {\RIGHTARROW} \kw{Real}
\ep
}
\pos{\end{multicols}}{}

\mnewfoil

\noindent
\begynd
\pind One can not add temperatures -- makes no sense in physics\,!\label{soosav-temp}
\begynd
\pind But one can take the mean value of two (or more) temperatures.
\pind One can subtract temperatures \nyl obtaining positive or negative
      temperature intervals.
\pind One can take the mean of any number of temperature, \nyl
      but would probably be well advised to have these represent \nyl
      regular sampling, or at least time-stamped.
\pind One can also define \sfsl{rate of change of temperature}.
\afslut
\mnewfoil
\bp
\kw{type}\\
\>\>Temp, MeanTemp, Degrees, TempIntv {\EQ} Degrees\\
\kw{value}\\
\>\>mean: Temp\kw{-set} {\TIMES} \kw{Nat} {\RIGHTARROW} MeanTemp\\
\>\>{\MINUS}: Temp {\TIMES} Temp {\RIGHTARROW} TempIntv\\
\kw{type}\\
\>\>TST {\EQ} (Temp {\TIMES} $\mathbb{T}$)\kw{-set}\\
\kw{value}\\
\>\>avg: TST {\RIGHTARROW} MeanTemp \\
\kw{type}\\
\>\>TimeUnit {\EQ} {\LBRACE}{\BAR}''year'',''month'',''day'''',hour'',{\DOTDOTDOT}{\BAR}{\RBRACE}\ \ \ \ \\
\>\>RoTC {\EQ} TempIntv {\TIMES} TimeUnit
\ep
\pind Etcetera.
\mnewfoil
\pind We leave it to the \pos{reader}{listener} to speculate on \nyl which
      operations one can perform on a persons' attributes: \nyl height, weight,
      birth date, name, etc. 
\pind And similarly for other domains.
\pind It is time to ``lift'' these observations.
\begynd
\pind After the examples above we should inquire as to \nyl which kind of
      units we may operate upon.
\pind For the sake of our later exposition it is enough that we look
      \nyl
      in some detail at the ``universe'' of physics.
\afslut
\afslut

\nbbbb{Physics Attributes}\label{si}

\bbb{SI: The International System of Quantities}

\begynd
\pind In physics we operate on \nyl values of attributes of manifest,
      i.e., physical phenomena.
\pind The type of some of these attributes are recorded \nyl in well
      known tables, cf.\,Tables\,\ref{table:1}--\ref{table:3}. 
\afslut


\mnewfoil

Table\,\vref{table:1} shows the base units of physics.

\begin{table}[h]\LLLL\HHHH
  \centering
\begin{tabular}{|lll|} \hline
Base quantity     &          Name   &      Type \\ \hline
length            &          meter   &     m \\
mass              &          kilogram   &  kg \\
time              &          second  &     s \\
electric current     &       ampere   &    A \\
thermodynamic temperature &  kelvin  &     K \\
amount of substance    &     mole    &     mol \\
luminous intensity     &     candela &     cd \\ \hline
\end{tabular} \caption{\HHHH Base Units} \label{table:1}
\end{table}

\mnewfoil

Table\,\vref{table:2} shows the units of physics derived from the base units.\LLLL\LLll%

\begin{table}[h]\LLLL\LLll
  \centering
\begin{tabular}{|llll|} \hline
         Name   &      Type & Derived Quantity & Derived Type \\ \hline
         radian & rad & angle & m/m  \\ 
steradian & sr & solid angle &          m$^2\times$m$^{-2}$ \\  
         Hertz  & Hz  & frequency & s$^{-1}$ \\
         newton & N   & force, weight   &
         kg{$\times$}m{$\times$}s$^{-2}$ \\
         pascal & Pa & pressure, stress & N/m$^2$ \\
         joule  & J & energy, work, heat & N$\times$m \\
         watt   & W & power, radiant flux &     J/s \\
         coulomb & C & electric charge & s$\times$A \\
volt &  V & voltage, electromotive force &  W/A (kg$\times$m$^2\times$s$^{-3}\times$A$^{-1}$) \\
farad & F & capacitance & C/V
         (kg$^{-1}\times$·m$^{-2}\times$s$^4\times$A$^2$) \\
ohm & $\Omega$ &  electrical resistance & V/A
         (kg$\times$m$^2\times$s$^{−3}\times$A$^{−2}$) \\
siemens & S & electrical conductance & A/V (kg${−1}\times$m$^{−2}\times$s$^3\times$A$^2$\\
weber & Wb & magnetic flux &  V$\times$s  (kg$\times$m$^2\times$s$^{-2}\times$A$^{-1}$)\\
tesla   &       T      &  magnetic flux density   & Wb/m$^2$
         (kg$\times$s$^{−2}\times$A$^{-1}$) \\
henry   &       H     &   inductance    &   Wb/A
         (kg$\times$m$^2\times$s$^{-2}\times$A$^2$) \\
degree Celsius    &     $^o$C    &   temperature relative to 273.15 K &
         K \\
lumen &  lm  &   luminous flux   & cd$\times$sr   (cd) \\
lux   &  lx  & illuminance   &   lm/m$^2$   (m$^{−2}\times$cd) \\ \hline
\end{tabular}  \caption{\HHHH Derived Units} \label{table:2}
\end{table}

\mnewfoil

Table\,\vref{table:3} shows further units of physics derived from the base units.\LLLL\LLll%

\begin{table}[h]\LLLL\LLll
  \centering
\begin{tabular}{|lll|} \hline
Name   &  Explanation & Derived Type \\ \hline
area                               & square meter                &    m$^2$ \\
volume                            & cubic meter                 &    m$^3$ \\
speed, velocity                   & meter per second             &   m/s \\
acceleration                     & meter per second squared   &     m/s$^2$ \\
wave number                        & reciprocal meter            &    m-1 \\
mass density                       & kilogram per cubic meter     &   kg/m$^3$ \\
specific volume                     & cubic meter per kilogram    &    m3/kg \\
current density                    & ampere per square meter     &    A/m$^2$ \\
magnetic field strength           & ampere per meter            &    A/m \\
amount-of-substance concentration   & mole per cubic meter       &     mol/m$3$ \\
luminance                           & candela per square meter    &    cd/m$^2$ \\
mass fraction                       & kilogram per kilogram       &    kg/kg = 1 \\ \hline
\end{tabular}
  \caption{\HHHH Further Units}
  \label{table:3}
\end{table}

\mnewfoil

Table\,\vref{table:4} shows standard prefixes for SI units of measure.

\begin{table}[h]
  \centering
\LLLL
\begin{tabular}{|llllllllllll|} \hline
Prefix name   &      & deca  &  hecto &  kilo &   mega &   giga &   tera  &  peta  &  exa  &   zetta &  yotta\\
Prefix symbol &      & da    &  h  &     k   &    M   &    G  &     T  &     P &      E &      Z &     Y\\
Factor        & 10$^0$ & 10$^1$ & 10$^2$ & 10$^3$ & 10$^6$  & 10$^9$  & 10$^{12}$  & 10$^{15}$ & 10$^{18}$ & 10$^{21}$ & 10$^{24}$\\\hline
\end{tabular}
  \caption{\HHHH Standard Prefixes for SI Units of Measure}
  \label{table:4}
\end{table}


Table\,\vref{table:5} shows fractions of SI units of measure.

\begin{table}[h]\LLLL\LLll\pos{\small}{}
  \centering
\begin{tabular}{|llllllllllll|} \hline
Prefix name      &    &   deci  &  centi  & milli &  micro &  nano &   pico &   femto &  atto &   zepto  & yocto\\
Prefix symbol     &   &   d   &    c  &     m   &    $\mu$ &       n &      p &      f &      a &      z    &   y\\
Factor        & 10$^0$ & 10$^{-1}$ & 10$^{-2}$ & 10$^{-3}$ & 10$^{-6}$  & 10$^{-9}$  & 10$^{-12}$  & 10$^{-15}$ & 10$^{-18}$ & 10$^{-21}$ & 10$^{-24}$\\\hline
\end{tabular}
  \caption{\HHHH Fractions}
  \label{table:5}
\end{table}
\HHHH

\mnewfoil

These ``pictures'' are meant as an eye opener, a ``teaser''.

\begin{center}
\epsfig{file=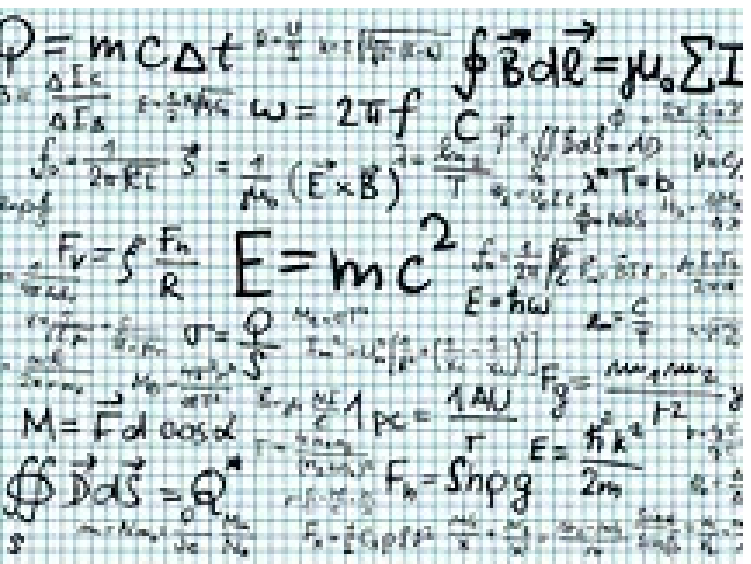,height=\pos{5}{9.2}cm}
\epsfig{file=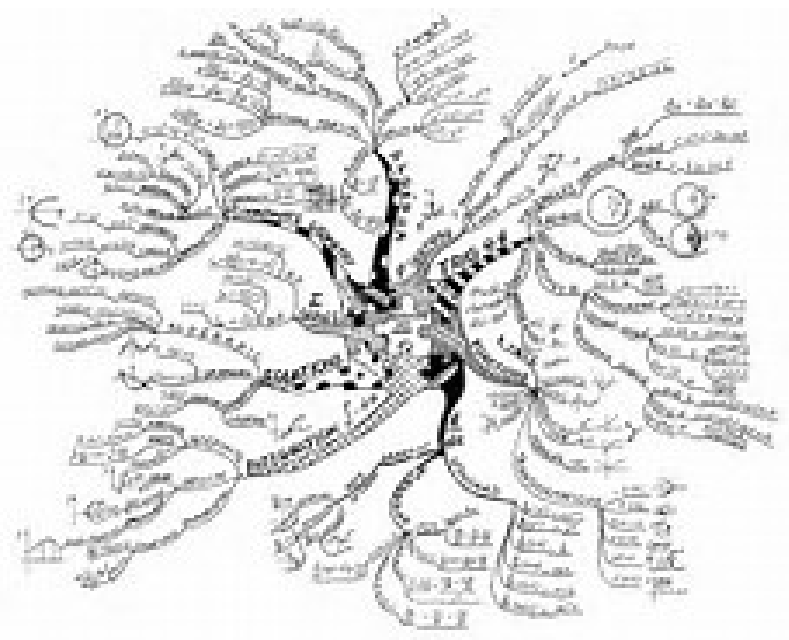,height=\pos{5}{9.2}cm}
\end{center}

\mnewfoil


And these formulas likewise\,!

\vspace{1mm}
 
\hrule
\begin{center}
\epsfig{file=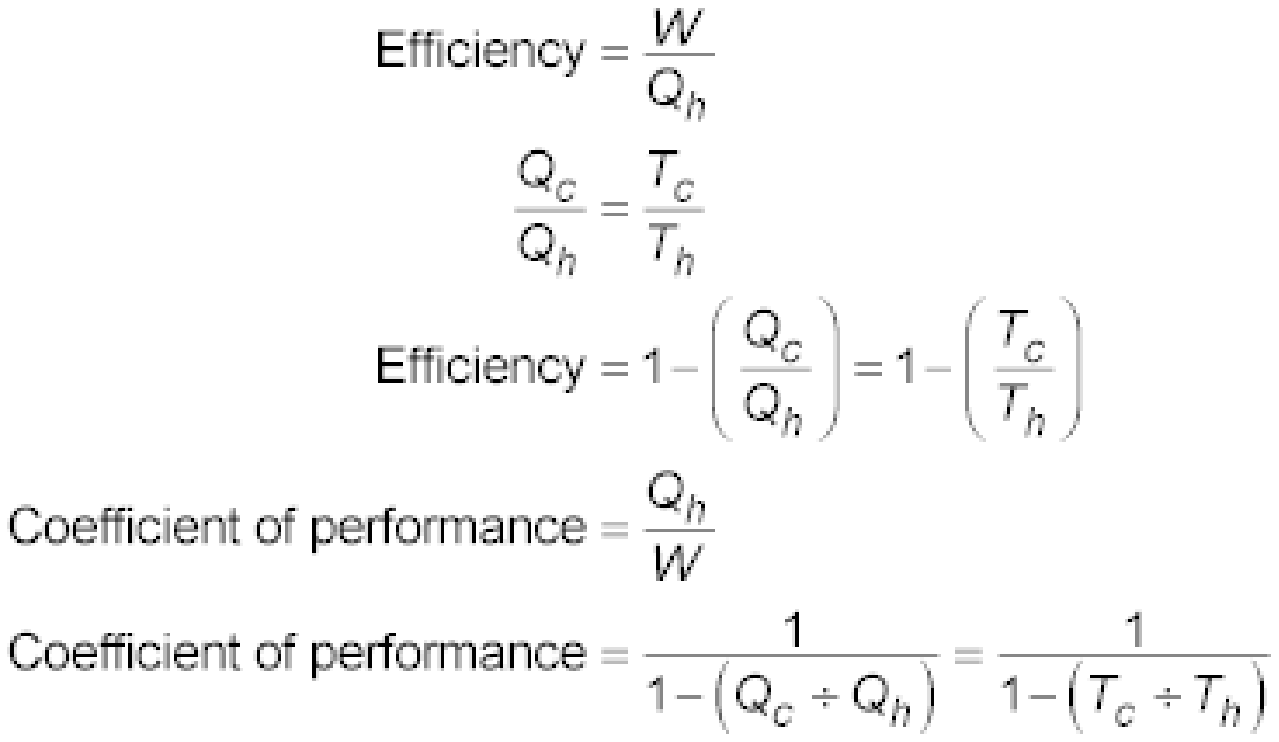,height=\pos{4}{8}cm}
\epsfig{file=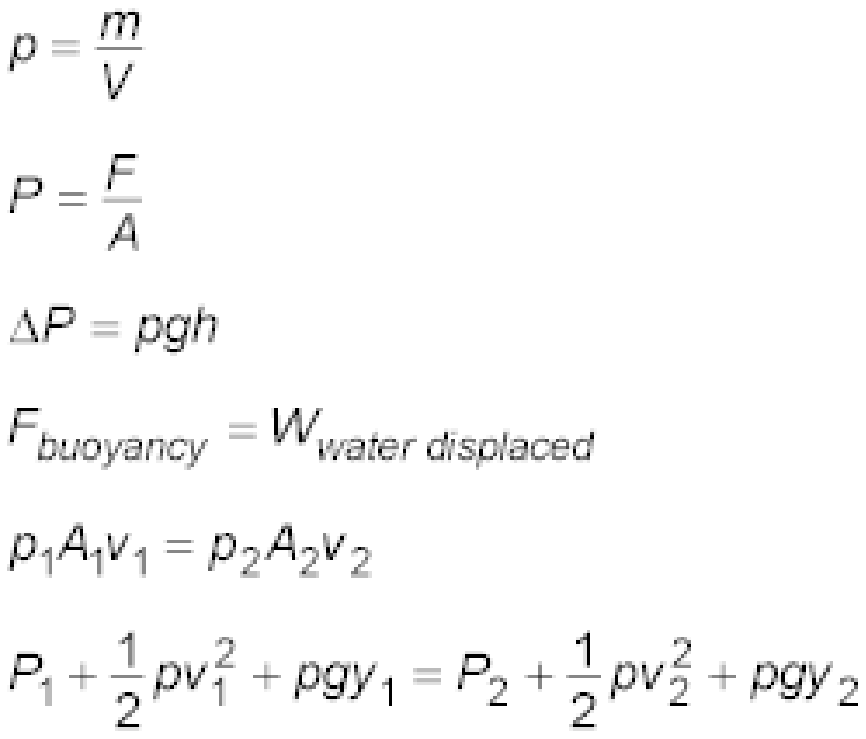,height=\pos{4}{8}cm}
\hrule
\vspace{1mm}
 
Carnot Engine \hfill Bernoulli Flow
\end{center}

\begynd
\pind The point in bringing this material is
\begynd
\pind that when modelling, i.e., describing domains
\pind we must be extremely careful in not falling into the trap
\pind of modelling physics, etc., types as we do in programming\,!
\afslut
\afslut

\nbbb{What Are We to Learn from this Exposition\,?}

\begynd
\pind \pos{We see from the previous section \pos{, Sect.\,\ref{si},}{} that
      p}{P}hysics units can be highly ``structured''\footnote{\LLLL For
        example, \textsf{Newton: kg{$\times$}m{$\times$}s$^{-2}$, Volt =
        kg$\times$m$^2\times$s$^{-3}\times$A$^{-1}$,} etc.}.
\pind What Are We to Learn from this Exposition\,?
\pind I think it is this:
\begynd
\pind It is customary, in programs of languages \nyl from \texttt{Algol
      60} via \texttt{Pascal} to \texttt{Java,} \nyl to assign
      \texttt{float} or \texttt{double}\footnote{\LLLL representing
        single-, resp.\ double-precision \textsf{32-bit IEEE 
      754 floating point} values} \textbf{type}s, as in
      \texttt{Java}, \nyl
      to [constants or] variables \nyl that for example represent values of physics.
\pind \sfsl{So rather completely different types of physics units \nyl are all
      cast into a same, simple-minded, ``number'' type.}
\pind \lilacolor{No chance, really, for any meaningful type checking.}
\afslut
\afslut

\nbbbb{Attribute Types, Scales and Values: Some Thoughts}\label{attrs:tsv}\label{attr:tsv}

\begynd
\pind This section further elaborates on the treatment of attributes given in
      Sect.\,\pos{\ref{Attributes}}{2.2.3},
      \pos{Pages}{Slides}\,\pageref{Attributes}--\pageref{Attributes.n}. 
\pind The elaboration is only sketched.
\pind It need be studied, in detail.
\mnewfoil

\pind The elaboration is this:
\begynd
\pind The \textsf{attr\_A} observer function, \nyl
      for a part $p$ of sort \textsf{P}, \nyl
      such as defined in Sect.\,\pos{\ref{Attributes}}{2.2.3}
        (\pos{Page}{Slide}\,\pageref{attr-A})\nyl
      yields values of type \textsf{A}.
\pind In the revised understanding of attributes\nyl
      the  \textsf{attr\_A} observer is now to yield \nyl both the type, \textsf{AT}, and
      the value, \textsf{AV}, of attribute A:     
\bp
\kw{type}\\
\>\>AT, AV\\
\kw{value}\\
\>\>attr\_A: P {\RIGHTARROW} AT {\TIMES} AV\ \ \ \ \ 
\ep
\pind You may think of \textsf{A} being defined by \textsf{AT {\TIMES} AV}.
\afslut
\afslut

\mnewfoil

\begynd
\pind The revision is further that a domain analysis \& description
      \nyl of the operations over attributes values, $\theta$:
\bp
\>\>\ $\theta$: A$_i${\TIMES}A$_j${\TIMES}{\DOTDOTDOT}{\TIMES}A$_k$ {\RIGHTARROW} V
\ep
be carefully checked -- such as hinted at in
      Sect.\,\pos{\vref{soosav}}{3.1
      (Slides\,\pageref{soosav-time}--\pageref{soosav-temp})}.
      
\pind Whether such operator-checks be researched and documented
\begynd
\pind ``once-and-for-all'' for given ``standard'' domains, by
      domain scientists, or
\pind per domain model, by domain engineers, \nyl
      in connection with specific software development projects
\afslut is left for you to decide\,!
\pind These operator-checks, together with an otherwise appropriate
      domain analysis \& description, 
\begynd
\pind \bbcolor{if not pursued, results in implicit semantics, and
\pind if pursued, results in explicit semantics.}
\afslut
\pind It is as simple as that\,!
\afslut


\nbbbbb{Conclusion}

\bbbb{What Have We Achieved\,?}

\begynd
\pind We have suggested that the issue of implicit semantics
      \cite{impex-0} be resolved
\begynd
\pind by providing a carefully analysed and described domain model
      \cite{BjornerFAoC2015MDAAD} 
      \nyl prior to requirements capture and software design,
\pind a both informally annotated and formally specified model \nyl
      that goes beyond \cite{BjornerFAoC2015MDAAD} in its treatment of
      attributes 
\pind in that these are now endowed with types [and possibly scales (or
      fractions)] and that each specific domain model analyses and
      formalises the constraints that operations upon attribute values
      are carefully analysed, statically.
\afslut
\afslut

\nbbbb{Domain Descriptions as Basis for Requirements Prescriptions}
\LLLL\HHHH
\begynd
\pind This \pos{paper}{invited talk} covers but one aspect of software
      development.
\afslut
\begin{itemize}
\item \cite{BjornerFAoCFacets} covers additional facets of domain
      analysis \& description.
\item {\cite{BjornerFAoC2015Req} offers a systematic approach
      \nyl to
      requirements engineering based on domain descriptions. \nyl
      It is this approach that justifies our claim that domain
      modelling}
      {\sfsl{``alleviate the issue of implicit semantics.''}}
\item \cite{BjornerFAoCProcesses} presents an operational/denotational
      semantics of the manifest domain analysis \& description
      calculus of \cite{BjornerFAoC2015MDAAD}.
\item \cite{BjornerMereologyCSP2017}\footnote{\LLLL Accepted for
            publication in \sfsl{Journal of Logical and Algebraic
            Methods in Programming}, 2018.} shows that to every
            manifest mereology there corresponds a \texttt{CSP} expression.
\item \cite{BjornerFAoCDemos} muses over issues of software
      simulators, demos, monitors and controllers.
\end{itemize}

\nbbbb{What Next\,?}

\begynd
\pind Well, there is a lot of fascinating research to be done now.
\pind Studying analysis \& description techniques for attribute types,
      values and constraints.
\pind And for engineering their support.
\afslut

\bbbb{Thanks}

\begynd
\pind to J. Paul Gibson and Dominique M{\'e}ry for inviting me,
\pind to J. Paul Gibson for organising my flights, hotel and
      registration, and
\pind to Dominique M{\'e}ry for his patience in waiting for my written contribution.
\afslut

\nbbbbb{Bibliographical Notes}

\bbbb{References to Draft Domain Descriptions}

\LLLL\HHHH
\begin{multicols}{2}
\begin{itemize}
\item \textsf{Swarms of Drones}\hfill\cite{BjornerDrones2017}
\item \textsf{\textsl{Urban Planning}} \hfill
  \cite{BjornerUrbanPlanningProcesses2017}
\item \textsf{\textsl{Documents}}  \hfill\cite{BjornerDocuments2017}
\item \textsf{\textsl{Credit Cards}}  \hfill\cite{BjornerCreditCard2016}
\item \textsf{\textsl{Weather \pos{Information}{} Systems}} \hfill
  \cite{BjornerWeather2016}
\item \textsf{\textsl{The Tokyo Stock Exchange}} \hfill \cite{db:tse:2010:www}
\item \textsf{\textsl{Pipelines}} \hfill \cite{2013pipe}
\item \textsf{\textsl{Road Transportation}} \hfill \cite{2013road}
\item \textsf{\textsl{Trans\pos{action}{.}-based Web Software}} \hfill
  \cite{evaKuhn2010}
\item \textsf{\textsl{``The Market''}} \hfill
  \cite{dines-kilov-02}
\item \textsf{\textsl{Container\pos{ [Shipping]}{} Lines}} \hfill
  \cite{db07:container}
\item \textsf{\textsl{Railway\pos{ System}{}s}} \hfill
  \cite{db00:p:ifac,dines-idpt-02,db03:ifac-cts2003,db02-amore-maint,db02-amore-ros} 
\end{itemize}
\end{multicols}

\bibliographystyle{eptcs}%

\renewcommand{\dbcite}[1]{\cite{#1}}\newcommand{\dbkey}[1]{}
\newcommand{\dbabstract}[1]{}\newcommand{\dbbibpth}{/home/db/bib}

\noindent
\begynd
\pind I apologise for the numerous references to own reports and publications.
\afslut

\normalsize\sf\small

\providecommand{\urlalt}[2]{\href{#1}{#2}}
\providecommand{\doi}[1]{doi:\urlalt{http://dx.doi.org/#1}{#1}}

\bibliography{%
\dbbibpth/strings,%
\dbbibpth/ada,%
\dbbibpth/alle,%
\dbbibpth/allrail,%
\dbbibpth/annals,%
\dbbibpth/arch,%
\dbbibpth/asm,%
\dbbibpth/asger-specification,%
\dbbibpth/asger-philosophy,%
\dbbibpth/b-method,%
\dbbibpth/bib,%
\dbbibpth/dsae,%
\dbbibpth/cofi,%
\dbbibpth/concept,%
\dbbibpth/darcsa,%
\dbbibpth/dcia,%
\dbbibpth/des,%
\dbbibpth/art-dines,%
\dbbibpth/rept-dines,%
\dbbibpth/book-dines,
\dbbibpth/adm-dines,%
\dbbibpth/notes-dines,%
\dbbibpth/ckm,%
\dbbibpth/dines99,%
\dbbibpth/domain,%
\dbbibpth/doreso,%
\dbbibpth/filosofi,%
\dbbibpth/icse,%
\dbbibpth/internet,%
\dbbibpth/jackson,%
\dbbibpth/languages,%
\dbbibpth/larch,%
\dbbibpth/ll,%
\dbbibpth/logics,%
\dbbibpth/method,%
\dbbibpth/1999,%
\dbbibpth/ontology,%
\dbbibpth/pe,%
\dbbibpth/pipelines,%
\dbbibpth/raisepap,%
\dbbibpth/s2000,%
\dbbibpth/s2000-1,%
\dbbibpth/seeduc,%
\dbbibpth/tlt,%
\dbbibpth/unuiist,%
\dbbibpth/vdm,%
\dbbibpth/vdmfme,%
\dbbibpth/zall}

\providecommand{\CoFI}{CoFI} \providecommand{\CASL}{CASL}
  \providecommand{\cofiWWW}{http://www.brics.dk/Projects/CoFI}
  \providecommand{\cofiWWWpath}{http://www.brics.dk/Projects/CoFI}
  \providecommand{\cofiFTP}{ftp://ftp.brics.dk/Projects/CoFI}
  \providecommand{\url}[1]{#1} \providecommand{\href}[2]{#2}
  \providecommand{\cofiWWWref}[2]{\href{\cofiWWWpath/#1}{#2}}
  \providecommand{\cofiSN}[1] {\cofiWWWref{StudyNotes/#1.html}{Study Note~#1}}
  \providecommand{\cofiLDSN}[1] {\cofiWWWref{StudyNotes/Lang/#1.html}{Language
  Design Study Note~#1}} \providecommand{\cofiLDSNnoWWW}[1]
  {\href{\cofiFTP/StudyNotes/Lang/}{Language Design Study Note~#1}}
  \providecommand{\cofiNote}[1] {\cofiWWWref{Notes/#1/index.html}{Note~#1}}
  \providecommand{\cofiDissentNote}[1]
  {\cofiWWWref{Notes/#1/index.html}{Note~#1}} \providecommand{\cofiDocument}[1]
  {\cofiWWWref{Documents/#1/index.html}{Documents/#1}}
  \providecommand{\cofiTentativeDocument}[1]
  {\cofiWWWref{Documents/Tentative/#1/index.html}{Documents/Tentative/#1}}
\begin{thebibliography}{10}
\providecommand{\bibitemdeclare}[2]{}
\providecommand{\surnamestart}{}
\providecommand{\surnameend}{}
\providecommand{\urlprefix}{Available at }
\providecommand{\url}[1]{\texttt{#1}}
\providecommand{\href}[2]{\texttt{#2}}
\providecommand{\urlalt}[2]{\href{#1}{#2}}
\providecommand{\doi}[1]{doi:\urlalt{http://dx.doi.org/#1}{#1}}
\providecommand{\bibinfo}[2]{#2}

\bibitemdeclare{inproceedings}{impex-0}
\bibitem{impex-0}
\bibinfo{author}{Yamine~A{\"{\i}}t \surnamestart Ameur\surnameend},
  \bibinfo{author}{J.~Paul \surnamestart Gibson\surnameend} \&
  \bibinfo{author}{Dominique \surnamestart M{\'{e}}ry\surnameend}
  (\bibinfo{year}{2014}): \emph{\bibinfo{title}{On Implicit and Explicit
  Semantics: Integration Issues in Proof-Based Development of Systems}}.
\newblock In: {\sl \bibinfo{booktitle}{Leveraging Applications of Formal
  Methods, Verification and Validation. Specialized Techniques and Applications
  - 6th International Symposium, ISoLA 2014, Imperial, Corfu, Greece, October
  8-11, 2014, Proceedings, Part {II}}}, pp. \bibinfo{pages}{604--618},
  \doi{10.1007/978-3-662-45231-8\_50}.

\bibitemdeclare{article}{impex-1}
\bibitem{impex-1}
\bibinfo{author}{Yamine~A{\"{\i}}t \surnamestart Ameur\surnameend} \&
  \bibinfo{author}{Dominique \surnamestart M{\'{e}}ry\surnameend}
  (\bibinfo{year}{2016}): \emph{\bibinfo{title}{Making explicit domain
  knowledge in formal system development}}.
\newblock {\sl \bibinfo{journal}{Sci. Comput. Program.}} \bibinfo{volume}{121},
  pp. \bibinfo{pages}{100--127}, \doi{10.1016/j.scico.2015.12.004}.

\bibitemdeclare{inproceedings}{db00:p:ifac}
\bibitem{db00:p:ifac}
\bibinfo{author}{Dines \surnamestart Bj{\o}rner\surnameend}
  (\bibinfo{year}{2000}): \emph{\bibinfo{title}{{Formal Software Techniques in
  Railway Systems}}}.
\newblock In \bibinfo{editor}{Eckehard \surnamestart Schnieder\surnameend},
  editor: {\sl \bibinfo{booktitle}{{9th IFAC Symposium on Control in
  Transportation Systems}}}, \bibinfo{organization}{{\sf VDI/VDE\--Gesellschaft
  Mess-- und Automatisieringstechnik}, {\sf VDI\--Gesellschaft f{\"u}r
  Fahrzeug-- und Verkehrstechnik}}, \bibinfo{address}{Technical University,
  Braunschweig, Germany}, pp. \bibinfo{pages}{1--12}.
\newblock \bibinfo{note}{Invited talk}.

\bibitemdeclare{inproceedings}{dines-kilov-02}
\bibitem{dines-kilov-02}
\bibinfo{author}{Dines \surnamestart Bj{\o}rner\surnameend}
  (\bibinfo{year}{2002}): \emph{\bibinfo{title}{{Domain Models of "The Market"
  --- in Preparation for E--Transaction Systems}}}.
\newblock In: {\sl \bibinfo{booktitle}{{Practical Foundations of Business and
  System Specifications (Eds.: Haim Kilov and Ken Baclawski)}}},
  \bibinfo{publisher}{Kluwer Academic Press}, \bibinfo{address}{The
  Netherlands}.
\newblock \bibinfo{note}{\htmladdnormallink{Final draft
  version}{http://www2.imm.dtu.dk/~db/themarket.pdf}}.

\bibitemdeclare{inproceedings}{db03:ifac-cts2003}
\bibitem{db03:ifac-cts2003}
\bibinfo{author}{Dines \surnamestart Bj{\o}rner\surnameend}
  (\bibinfo{year}{2003}): \emph{\bibinfo{title}{{Dynamics of Railway Nets: On
  an Interface between Automatic Control and Software Engineering}}}.
\newblock In: {\sl \bibinfo{booktitle}{{CTS2003: 10th IFAC Symposium on Control
  in Transportation Systems}}}, \bibinfo{publisher}{Elsevier Science Ltd.},
  \bibinfo{address}{Oxford, UK}, \doi{10.1016/S1474-6670(17)32424-2}.
\newblock \bibinfo{note}{Symposium held at Tokyo, Japan. Editors: S. Tsugawa
  and M. Aoki. \htmladdnormallink{Final
  version}{http://www2.imm.dtu.dk/~db/ifac-dynamics.pdf}}.

\bibitemdeclare{book}{TheSEBook2wo}
\bibitem{TheSEBook2wo}
\bibinfo{author}{Dines \surnamestart Bj{\o}rner\surnameend}
  (\bibinfo{year}{2006}): \emph{\bibinfo{title}{{Software Engineering, Vol.~2:
  Specification of Systems and Languages}}}.
\newblock \bibinfo{series}{Texts in Theoretical Computer Science, the EATCS
  Series}, \bibinfo{publisher}{Springer}.
\newblock \bibinfo{note}{{Chapters 12--14 are primarily authored by Christian
  Krog Madsen.}}

\bibitemdeclare{techreport}{db07:container}
\bibitem{db07:container}
\bibinfo{author}{Dines \surnamestart Bj{\o}rner\surnameend}
  (\bibinfo{year}{2007}): \emph{\bibinfo{title}{{A Container Line Industry
  Domain}}}.
\newblock \bibinfo{type}{Techn. Report}, \bibinfo{address}{Fredsvej 11, DK-2840
  Holte, Denmark}.
\newblock \bibinfo{note}{\htmladdnormallink{Extensive
  Draft}{http://www2.imm.dtu.dk/~db/container-paper.pdf}}.

\bibitemdeclare{inproceedings}{dines:ugo65:2008}
\bibitem{dines:ugo65:2008}
\bibinfo{author}{Dines \surnamestart Bj{\o}rner\surnameend}
  (\bibinfo{year}{2008}): \emph{\bibinfo{title}{{From Domains to
  Requirements}}}.
\newblock In: {\sl \bibinfo{booktitle}{{Montanari Festschrift}}}, {\sl
  \bibinfo{series}{Lecture Notes in Computer Science (eds.\ Pierpaolo Degano,
  Rocco De Nicola and Jos{\'e} Meseguer)}} \bibinfo{volume}{5065},
  \bibinfo{publisher}{Springer}, \bibinfo{address}{Heidelberg}, pp.
  \bibinfo{pages}{1--30}.

\bibitemdeclare{inproceedings}{dines:facs:2008}
\bibitem{dines:facs:2008}
\bibinfo{author}{Dines \surnamestart Bj{\o}rner\surnameend}
  (\bibinfo{year}{2010}): \emph{\bibinfo{title}{{Domain Engineering}}}.
\newblock In \bibinfo{editor}{Paul \surnamestart Boca\surnameend} \&
  \bibinfo{editor}{Jonathan \surnamestart Bowen\surnameend}, editors: {\sl
  \bibinfo{booktitle}{{Formal Methods: State of the Art and New Directions}}},
  \bibinfo{series}{{Eds.\ Paul Boca and Jonathan Bowen}},
  \bibinfo{publisher}{Springer}, \bibinfo{address}{London, UK}, pp.
  \bibinfo{pages}{1--42}, \doi{10.1007/978-1-84882-736-3\_1}.

\bibitemdeclare{techreport}{evaKuhn2010}
\bibitem{evaKuhn2010}
\bibinfo{author}{Dines \surnamestart Bj{\o}rner\surnameend}
  (\bibinfo{year}{2010}): \emph{\bibinfo{title}{{On Development of Web-based
  Software: A Divertimento of Ideas and Suggestions}}}.
\newblock \bibinfo{type}{{Technical}}, \bibinfo{institution}{{Technical
  University of Vienna}}.
\newblock \bibinfo{note}{Http://www.imm.dtu.dk/\~{}dibj/wfdftp.pdf}.

\bibitemdeclare{incollection}{dines-maurer}
\bibitem{dines-maurer}
\bibinfo{author}{Dines \surnamestart Bj{\o}rner\surnameend}
  (\bibinfo{year}{2011}): \emph{\bibinfo{title}{{Domains: Their Simulation,
  Monitoring and Control -- A Divertimento of Ideas and Suggestions}}}.
\newblock In: {\sl \bibinfo{booktitle}{{Rainbow of Computer Science,
  Festschrift for Hermann Maurer on the Occasion of His 70th Anniversary.}}},
  \bibinfo{series}{Festschrift (eds.\ C.~Calude, G.~Rozenberg and A.~Saloma)},
  \bibinfo{publisher}{Springer}, \bibinfo{address}{Heidelberg, Germany}, pp.
  \bibinfo{pages}{{167--183}}.

\bibitemdeclare{techreport}{2013pipe}
\bibitem{2013pipe}
\bibinfo{author}{Dines \surnamestart Bj{\o}rner\surnameend}
  (\bibinfo{year}{2013}):
  \emph{\bibinfo{title}{{\htmladdnormallinkfoot{Pipelines -- a Domain
  Description}{http:\-//\-www.\-imm.\-dtu.\-dk/\-\latexhtml{\~{}}{~}dibj/\-pipe-p.pdf}}}}.
\newblock \bibinfo{type}{{Experimental Research Report}}
  \bibinfo{number}{{2013-2}}, \bibinfo{institution}{{DTU Compute and Fredsvej
  11, DK-2840 Holte, Denmark}}.

\bibitemdeclare{techreport}{2013road}
\bibitem{2013road}
\bibinfo{author}{Dines \surnamestart Bj{\o}rner\surnameend}
  (\bibinfo{year}{2013}): \emph{\bibinfo{title}{{\htmladdnormallinkfoot{Road
  Transportation -- a Domain
  Description}{http://www.imm.dtu.dk/\latexhtml{\~{}}{~}dibj/road-p.pdf}}}}.
\newblock \bibinfo{type}{{Experimental Research Report}}
  \bibinfo{number}{{2013-4}}, \bibinfo{institution}{{DTU Compute and Fredsvej
  11, DK-2840 Holte, Denmark}}.

\bibitemdeclare{incollection}{2013da-jaist}
\bibitem{2013da-jaist}
\bibinfo{author}{Dines \surnamestart Bj{\o}rner\surnameend}
  (\bibinfo{year}{2014}): \emph{\bibinfo{title}{{Domain Analysis: Endurants --
  An Analysis \& Description Process Model}}}.
\newblock In \bibinfo{editor}{{Shusaku} \surnamestart Iida\surnameend},
  \bibinfo{editor}{{Jos{\'e}} \surnamestart Meseguer\surnameend} \&
  \bibinfo{editor}{{Kazuhiro} \surnamestart Ogata\surnameend}, editors: {\sl
  \bibinfo{booktitle}{{Specification, Algebra, and Software: A Festschrift
  Symposium in Honor of Kokichi Futatsugi}}}, \bibinfo{publisher}{Springer}.

\bibitemdeclare{techreport}{BjornerCreditCard2016}
\bibitem{BjornerCreditCard2016}
\bibinfo{author}{Dines \surnamestart Bj{\o}rner\surnameend}
  (\bibinfo{year}{2016}): \emph{\bibinfo{title}{{A Credit Card System: Uppsala
  Draft}}}.
\newblock \bibinfo{type}{{Technical Report: Experimental Research}},
  \bibinfo{institution}{{Fredsvej 11, DK--2840 Holte, Denmark}}.
\newblock \bibinfo{note}{Http://www.imm.dtu.dk/\~{}dibj/2016/credit/accs.pdf}.

\bibitemdeclare{article}{BjornerFAoCProcesses}
\bibitem{BjornerFAoCProcesses}
\bibinfo{author}{Dines \surnamestart Bj{\o}rner\surnameend}
  (\bibinfo{year}{2016}): \emph{\bibinfo{title}{{Domain Analysis and
  Description -- Formal Models of Processes and Prompts}}}.
\newblock \bibinfo{note}{Extensive revision of \cite{2013da-jaist}.
  http://\-www.\-imm.\-dtu.\-dk\-/\~{}dibj/\-2016/\-process/\-process-p.pdf}.

\bibitemdeclare{article}{BjornerFAoCFacets}
\bibitem{BjornerFAoCFacets}
\bibinfo{author}{Dines \surnamestart Bj{\o}rner\surnameend}
  (\bibinfo{year}{2016}): \emph{\bibinfo{title}{{Domain Facets: Analysis \&
  Description}}}.
\newblock \bibinfo{note}{Extensive revision of \cite{dines:facs:2008}.
  http://\-www.\-imm.\-dtu.\-dk\-/\~{}dibj/\-2016/\-facets/\-faoc-facets.pdf}.

\bibitemdeclare{techreport}{BjornerFAoCDemos}
\bibitem{BjornerFAoCDemos}
\bibinfo{author}{Dines \surnamestart Bj{\o}rner\surnameend}
  (\bibinfo{year}{2016}): \emph{\bibinfo{title}{{Domains: Their Simulation,
  Monitoring and Control -- A Divertimento of Ideas and Suggestions}}}.
\newblock \bibinfo{type}{Technical Report}, \bibinfo{institution}{{Fredsvej 11,
  DK--2840 Holte, Denmark}}.
\newblock \bibinfo{note}{Extensive revision of \cite{dines-maurer}.
  http://\-www.\-imm.\-dtu.\-dk\-/\~{}di\-bj/\-2016\-/demos/\-faoc-\-demo.pdf}.

\bibitemdeclare{article}{BjornerFAoC2015Req}
\bibitem{BjornerFAoC2015Req}
\bibinfo{author}{Dines \surnamestart Bj{\o}rner\surnameend}
  (\bibinfo{year}{2016}): \emph{\bibinfo{title}{{From Domain Descriptions to
  Requirements Prescriptions -- A Different Approach to Requirements
  Engineering}}}.
\newblock \bibinfo{note}{{Extensive revision of \cite{dines:ugo65:2008}}}.

\bibitemdeclare{techreport}{BjornerWeather2016}
\bibitem{BjornerWeather2016}
\bibinfo{author}{Dines \surnamestart Bj{\o}rner\surnameend}
  (\bibinfo{year}{2016}): \emph{\bibinfo{title}{{Weather Information Systems:
  Towards a Domain Description}}}.
\newblock \bibinfo{type}{{Technical Report: Experimental Research}},
  \bibinfo{institution}{{Fredsvej 11, DK--2840 Holte, Denmark}}.
\newblock \bibinfo{note}{Http://www.imm.dtu.dk/\~{}dibj/2016/wis/wis-p.pdf}.

\bibitemdeclare{techreport}{BjornerDrones2017}
\bibitem{BjornerDrones2017}
\bibinfo{author}{Dines \surnamestart Bj{\o}rner\surnameend}
  (\bibinfo{year}{2017}): \emph{\bibinfo{title}{{A Space of Swarms of
  Drones}}}.
\newblock \bibinfo{type}{{Research Note}}.
\newblock
  \bibinfo{note}{Http://\-www.\-imm.\-dtu.\-dk/\-\~{}dibj/\-2017/\-docs/\-docs.pdf}.

\bibitemdeclare{techreport}{BjornerDocuments2017}
\bibitem{BjornerDocuments2017}
\bibinfo{author}{Dines \surnamestart Bj{\o}rner\surnameend}
  (\bibinfo{year}{2017}): \emph{\bibinfo{title}{{What are Documents\,?}}}
\newblock \bibinfo{type}{{Research Note}}.
\newblock
  \bibinfo{note}{Http://\-www.\-imm.\-dtu.\-dk/\-\~{}dibj/\-2017/\-docs/\-docs.pdf}.

\bibitemdeclare{techreport}{2018:Bjorner:philo}
\bibitem{2018:Bjorner:philo}
\bibinfo{author}{Dines \surnamestart Bj{\o}rner\surnameend}
  (\bibinfo{year}{2018}): \emph{\bibinfo{title}{{A Philosophy of Domain Science
  \& Engineering -- An Interpretation of Kai S{\o}rlander's Philosophy}}}.
\newblock \bibinfo{type}{{Research Note}}.
\newblock
  \bibinfo{note}{{http://\-www.\-imm.\-dtu.\-dk/\~{}dibj/\-2018/\-philosophy/\-filo.pdf}}.

\bibitemdeclare{article}{BjornerMereologyCSP2017}
\bibitem{BjornerMereologyCSP2017}
\bibinfo{author}{Dines \surnamestart Bj{\o}rner\surnameend}
  (\bibinfo{year}{{2018}}): \emph{\bibinfo{title}{{To Every Manifest Domain a
  \texttt{CSP} Expression --- A R{\^o}le for Mereology in Computer Science}}}.
\newblock {\sl \bibinfo{journal}{{Journal of Logical and Algebraic Methods in
  Programming}}} (\bibinfo{number}{94}), pp. \bibinfo{pages}{91--108},
  \doi{10.1016/j.jlamp.2017.09.005}.

\bibitemdeclare{techreport}{db:tse:2010:www}
\bibitem{db:tse:2010:www}
\bibinfo{author}{Dines \surnamestart Bj{\o}rner\surnameend}
  (\bibinfo{year}{January and February, 2010}): \emph{\bibinfo{title}{{The
  Tokyo Stock Exchange Trading Rules}}}.
\newblock \bibinfo{type}{{R\&D Experiment}}, \bibinfo{address}{{Fredsvej 11,
  DK-2840 Holte, Denmark}}.
\newblock \bibinfo{note}{\htmladdnormallink{Version
  1}{http://www2.imm.dtu.dk/~db/todai/tse-1.pdf}, \htmladdnormallink{Version
  2}{http://www2.imm.dtu.dk/~db/todai/tse-2.pdf}}.

\bibitemdeclare{article}{BjornerFAoC2015MDAAD}
\bibitem{BjornerFAoC2015MDAAD}
\bibinfo{author}{Dines \surnamestart Bj{\o}rner\surnameend}
  (\bibinfo{year}{Online: July 2016, Journal: March 2017}):
  \emph{\bibinfo{title}{{Manifest Domains: Analysis \& Description}}}.
\newblock {\sl \bibinfo{journal}{{Formal Aspects of Computing}}}
  \bibinfo{volume}{29}(\bibinfo{number}{2}), pp. \bibinfo{pages}{175--225},
  \doi{10.1007/s00165-016-0385-z}.

\bibitemdeclare{inproceedings}{dines-idpt-02}
\bibitem{dines-idpt-02}
\bibinfo{author}{Dines \surnamestart Bj{\o}rner\surnameend},
  \bibinfo{author}{Chris~W. \surnamestart George\surnameend} \&
  \bibinfo{author}{S{\o}ren \surnamestart Prehn\surnameend}
  (\bibinfo{year}{2002}): \emph{\bibinfo{title}{{Computing Systems for Railways
  --- A R{\^o}le for Domain Engineering. Relations to Requirements Engineering
  and Software for Control Applications}}}.
\newblock In: {\sl \bibinfo{booktitle}{{Integrated Design and Process
  Technology. Editors: Bernd Kraemer and John C. Petterson}}},
  \bibinfo{publisher}{Society for Design and Process Science},
  \bibinfo{address}{P.O.Box 1299, Grand View, Texas 76050-1299, USA}.
\newblock \bibinfo{note}{\htmladdnormallink{Extended
  version}{http://www2.imm.dtu.dk/~db/pasadena-25.pdf}}.

\bibitemdeclare{book}{e:db:Bj80f}
\bibitem{e:db:Bj80f}
\bibinfo{editor}{Dines \surnamestart Bj{\o}rner\surnameend} \&
  \bibinfo{editor}{Ole~N. \surnamestart Oest\surnameend}, editors
  (\bibinfo{year}{1980}): \emph{\bibinfo{title}{{Towards a Formal Description
  of {A}da}}}.
\newblock {\sl \bibinfo{series}{LNCS}}~\bibinfo{volume}{98},
  \bibinfo{publisher}{Springer}.

\bibitemdeclare{techreport}{BjornerUrbanPlanningProcesses2017}
\bibitem{BjornerUrbanPlanningProcesses2017}
\bibinfo{author}{Dines \surnamestart
  Bj{\o}rner\label{BjornerUrbanPlanningProcesses2017}\surnameend}
  (\bibinfo{year}{2017}): \emph{\bibinfo{title}{{Urban Planning Processes}}}.
\newblock \bibinfo{type}{{Research Note}}.
\newblock
  \bibinfo{note}{Http://\-www.\-imm.\-dtu.\-dk/\-\~{}dibj/\-2017/\-up/\-urban\--planning.pdf}.

\bibitemdeclare{techreport}{vdm:CCITT80}
\bibitem{vdm:CCITT80}
\bibinfo{author}{\surnamestart C.C.I.T.T.\surnameend} (\bibinfo{year}{1980}):
  \emph{\bibinfo{title}{{The Specification of CHILL}}}.
\newblock \bibinfo{type}{Technical Report} \bibinfo{number}{Recommendation
  Z200}, \bibinfo{institution}{International Telegraph and Telephone
  Consultative Committee}, \bibinfo{address}{Geneva, Switzerland}.

\bibitemdeclare{inproceedings}{Clem84}
\bibitem{Clem84}
\bibinfo{author}{G.B. \surnamestart Clemmensen\surnameend} \&
  \bibinfo{author}{O.~\surnamestart Oest\surnameend} (\bibinfo{year}{1984}):
  \emph{\bibinfo{title}{Formal Specification and Development of an {A}da
  Compiler -- A {VDM} Case Study}}.
\newblock In: {\sl \bibinfo{booktitle}{Proc. 7th International Conf. on
  Software Engineering, 26.-29. March 1984, Orlando, Florida}},
  \bibinfo{organization}{IEEE}, pp. \bibinfo{pages}{430--440}.

\bibitemdeclare{misc}{FDR2:2004}
\bibitem{FDR2:2004}
\bibinfo{author}{\surnamestart {Computing Laboratory, University of Oxford,
  England}\surnameend} (\bibinfo{year}{2003}): \emph{\bibinfo{title}{{FDR4: The
  CSP Refinement Checker}}}.
\newblock \bibinfo{howpublished}{Published on the Internet:
  \texttt{https://www.cs.ox.ac.uk/projects/fdr/}}.

\bibitemdeclare{book}{RSL}
\bibitem{RSL}
\bibinfo{author}{Chris~W. \surnamestart George\surnameend},
  \bibinfo{author}{Peter \surnamestart Haff\surnameend}, \bibinfo{author}{Klaus
  \surnamestart Havelund\surnameend}, \bibinfo{author}{Anne~Elisabeth
  \surnamestart Haxthausen\surnameend}, \bibinfo{author}{Robert \surnamestart
  Milne\surnameend}, \bibinfo{author}{Claus~Bendix \surnamestart
  Nielsen\surnameend}, \bibinfo{author}{S{\o}ren \surnamestart
  Prehn\surnameend} \& \bibinfo{author}{Kim~Ritter \surnamestart
  Wagner\surnameend} (\bibinfo{year}{1992}): \emph{\bibinfo{title}{The RAISE
  Specification Language}}.
\newblock \bibinfo{series}{The BCS Practitioner Series},
  \bibinfo{publisher}{Prentice-Hall}, \bibinfo{address}{Hemel Hampstead,
  England}.

\bibitemdeclare{incollection}{Haff87}
\bibitem{Haff87}
\bibinfo{author}{P.~\surnamestart Haff\surnameend} \& \bibinfo{author}{A.V.
  \surnamestart Olsen\surnameend} (\bibinfo{year}{1987}):
  \emph{\bibinfo{title}{Use of {VDM} within {CCITT}}}.
\newblock In: {\sl \bibinfo{booktitle}{{VDM -- A Formal Method at Work, eds.
  Dines Bj{\o}rner, Cliff B. Jones, Micheal Mac an Airchinnigh and Erich J.
  Neuhold}}}, \bibinfo{publisher}{Springer, Lecture Notes in Computer Science,
  Vol. 252}, pp. \bibinfo{pages}{324--330}.
\newblock \bibinfo{note}{{Proc. VDM-Europe Symposium 1987, Brussels, Belgium}}.

\bibitemdeclare{article}{Hoa78a}
\bibitem{Hoa78a}
\bibinfo{author}{{C.A.R.} \surnamestart Hoare\surnameend}
  (\bibinfo{year}{1978}): \emph{\bibinfo{title}{{Communicating Sequential
  Processes}}}.
\newblock {\sl \bibinfo{journal}{Communications of the ACM}}
  \bibinfo{volume}{21}(\bibinfo{number}{8}).

\bibitemdeclare{book}{Hoare85+2004}
\bibitem{Hoare85+2004}
\bibinfo{author}{{C.A.R.} \surnamestart Hoare\surnameend}
  (\bibinfo{year}{1985}): \emph{\bibinfo{title}{{Communicating Sequential
  Processes}}}.
\newblock \bibinfo{series}{{C.A.R.} Hoare Series in Computer Science},
  \bibinfo{publisher}{Prentice-Hall International}.
\newblock \bibinfo{note}{Published electronically:
  {http://www.\-using\-csp.\-com/\-csp\-book.pdf} (2004)}.

\bibitemdeclare{book}{Hoare85}
\bibitem{Hoare85}
\bibinfo{author}{{C.A.R.} \surnamestart Hoare\surnameend}
  (\bibinfo{year}{1985}): \emph{\bibinfo{title}{{Communicating Sequential
  Processes}}}.
\newblock \bibinfo{series}{{C.A.R.} Hoare Series in Computer Science},
  \bibinfo{publisher}{Prentice-Hall International}.

\bibitemdeclare{misc}{CARH:Electronic}
\bibitem{CARH:Electronic}
\bibinfo{author}{{C.A.R.} \surnamestart Hoare\surnameend}
  (\bibinfo{year}{2004}): \emph{\bibinfo{title}{{Communicating Sequential
  Processes}}}.
\newblock \bibinfo{howpublished}{Published electronically:
  \texttt{http://www.\-usingcsp.\-com/\-cspbook.pdf}}.
\newblock \bibinfo{note}{Second edition of \cite{Hoare85}. See also
  \texttt{http://\-www.\-usingcsp.\-com/}}.

\bibitemdeclare{book}{lexicon}
\bibitem{lexicon}
\bibinfo{author}{Michael~A. \surnamestart Jackson\surnameend}
  (\bibinfo{year}{1995}): \emph{\bibinfo{title}{{Software {R}equirements \&
  {S}pecifications: a lexicon of practice, principles and prejudices}}}.
\newblock \bibinfo{series}{ACM Press}, \bibinfo{publisher}{Addison-Wesley},
  \bibinfo{address}{Reading, England}.

\bibitemdeclare{book}{OED}
\bibitem{OED}
\bibinfo{author}{W.~\surnamestart Little\surnameend}, \bibinfo{author}{H.W.
  \surnamestart Fowler\surnameend}, \bibinfo{author}{J.~\surnamestart
  Coulson\surnameend} \& \bibinfo{author}{C.T. \surnamestart Onions\surnameend}
  (\bibinfo{year}{1973, 1987}): \emph{\bibinfo{title}{{The Shorter {O}xford
  English Dictionary on Historical Principles}}}.
\newblock \bibinfo{publisher}{Clarendon Press, Oxford, England}.
\newblock \bibinfo{note}{{Two vols.}}

\bibitemdeclare{inproceedings}{Oest86}
\bibitem{Oest86}
\bibinfo{author}{Ole~N. \surnamestart Oest\surnameend} (\bibinfo{year}{1986}):
  \emph{\bibinfo{title}{{VDM} From Research to Practice (Invited Paper)}}.
\newblock In: {\sl \bibinfo{booktitle}{{IFIP} Congress}}, pp.
  \bibinfo{pages}{527--534}.

\bibitemdeclare{inproceedings}{db02-amore-maint}
\bibitem{db02-amore-maint}
\bibinfo{author}{Martin \surnamestart P\v{e}ni\v{c}ka\surnameend},
  \bibinfo{author}{Albena~Kirilova \surnamestart Strupchanska\surnameend} \&
  \bibinfo{author}{Dines \surnamestart Bj{\o}rner\surnameend}
  (\bibinfo{year}{2003}): \emph{\bibinfo{title}{{Train Maintenance Routing}}}.
\newblock In: {\sl \bibinfo{booktitle}{{FORMS'2003: Symposium on Formal Methods
  for Railway Operation and Control Systems}}}, \bibinfo{publisher}{L'Harmattan
  Hongrie}.
\newblock \bibinfo{note}{Conf. held at Techn.Univ. of Budapest, Hungary.
  Editors: G. Tarnai and E. Schnieder, Germany. \htmladdnormallink{Final
  version}{http://www2.imm.dtu.dk/~db/martin.pdf}}.

\bibitemdeclare{inbook}{RoscoeFDR94a}
\bibitem{RoscoeFDR94a}
\bibinfo{author}{A.~W. \surnamestart Roscoe\surnameend} (\bibinfo{year}{1994}):
  \emph{\bibinfo{title}{Model checking CSP}}, pp. \bibinfo{pages}{353--378}.
\newblock \bibinfo{publisher}{Prentice\--Hall Intl.}

\bibitemdeclare{book}{Roscoe97}
\bibitem{Roscoe97}
\bibinfo{author}{A.~W. \surnamestart Roscoe\surnameend} (\bibinfo{year}{1997}):
  \emph{\bibinfo{title}{{Theory and Practice of Concurrency}}}.
\newblock \bibinfo{series}{{C.A.R.} Hoare Series in Computer Science},
  \bibinfo{publisher}{Prentice-Hall}.
\newblock \bibinfo{note}{Now available on the net:
  http://\-www.\-com\-lab.\-ox.\-ac.\-uk/\-people/\-bill.\-ros\-coe/pub\-li\-ca\-tions/\-68b.pdf}.

\bibitemdeclare{book}{Schneider99}
\bibitem{Schneider99}
\bibinfo{author}{Steve \surnamestart Schneider\surnameend}
  (\bibinfo{year}{2000}): \emph{\bibinfo{title}{{Concurrent and Real-time
  Systems --- The CSP Approach}}}.
\newblock \bibinfo{series}{Worldwide Series in Computer Science},
  \bibinfo{publisher}{John Wiley \& Sons, Ltd.}, \bibinfo{address}{Baffins
  Lane, Chichester, West Sussex PO19 1UD, England}.

\bibitemdeclare{inproceedings}{db02-amore-ros}
\bibitem{db02-amore-ros}
\bibinfo{author}{Albena~Kirilova \surnamestart Strupchanska\surnameend},
  \bibinfo{author}{Martin \surnamestart P\v{e}ni\v{c}ka\surnameend} \&
  \bibinfo{author}{Dines \surnamestart
  Bj{\o}rner\label{db02-amore-ros}\surnameend} (\bibinfo{year}{2003}):
  \emph{\bibinfo{title}{{Railway Staff Rostering}}}.
\newblock In: {\sl \bibinfo{booktitle}{{FORMS2003: Symposium on Formal Methods
  for Railway Operation and Control Systems}}}, \bibinfo{publisher}{L'Harmattan
  Hongrie}.
\newblock \bibinfo{note}{Conf. held at Techn.Univ. of Budapest, Hungary.
  Editors: G. Tarnai and E. Schnieder, Germany. \htmladdnormallink{Final
  version}{http://www2.imm.dtu.dk/~db/albena.pdf}}.

\end{thebibliography}

\end{document}